\documentclass[%
reprint,
 amsmath,amssymb,
 aps,
pra,
]{revtex4-2}
\usepackage{physics}
\usepackage{tikz}
\usetikzlibrary{decorations.markings, calc, arrows.meta, positioning, shadings}

\tikzset{laser/.style={thick, black}}
\tikzset{arrow inside/.style = {postaction=decorate, decoration={markings, mark=at position .62 with \arrow{stealth}}}}
\tikzset{arrow inside1/.style = {postaction=decorate, decoration={markings,
						 mark=at position .62 with \arrow{stealth}}}}
\tikzset{ray/.style={very thick, red, arrow inside}}
\tikzset{ray1/.style={very thick, red, arrow inside1}}
\tikzset{detector/.style={thick, draw=black, fill=black!40}}
\tikzset{reflector/.style={thick, black, left color=black!50, right color=black!50, middle color=white}}
\tikzset{reflector1/.style={thick, black, left color=black!50, right color=black!50, middle color=white}}

\newcommand{\ILE}[1]{\textcolor{magenta}{[{\tt ILE}: #1]}}
\usepackage{graphicx}
\usepackage{dcolumn}
\usepackage{bm}
\usepackage[colorlinks=true,linkcolor=blue,citecolor=blue,urlcolor=blue]{hyperref}

\usepackage{caption}
\usepackage{subcaption}

\usepackage{ragged2e}

\usepackage{bbm}
\usepackage{mathtools}
\usepackage{amsthm}
\usepackage{amssymb}
\usepackage{amsfonts}
\usepackage{amsmath}
\DeclareFontFamily{U}{mathx}{}
\DeclareFontShape{U}{mathx}{m}{n}{ <-> mathx10 }{}
\DeclareSymbolFont{mathx}{U}{mathx}{m}{n}
\DeclareFontSubstitution{U}{mathx}{m}{n}
\DeclareMathAccent{\widecheck}{0}{mathx}{"71}
\newcommand{\Id}{\mathbbm{I}} 
 
\DeclareMathOperator{\diff}{d} 

\renewcommand\vec{\mathbf}
\newcommand{\trans}{\mathsf{T}} 
\newcommand{\reals}{\mathbb{R}}
\newcommand{\complex}{\mathbb{C}}

\newcommand{\integers}{\mathbb{Z}}
\newcommand{\hilbert}{\mathcal{H}} 

\DeclareCaptionJustification{justified}{\justifying}

\begin{document}

\preprint{APS/123-QED}

\title{Squeezing-Enhanced Rotational Doppler Metrology}
\author{Javier Navarro}
\altaffiliation{jnavarro@bcamath.org}
\affiliation{Basque Center for Applied Mathematics (BCAM), Alameda de Mazarredo 14, 48009 Bilbao, Spain}
\affiliation{Department of Physical Chemistry, University of the Basque Country UPV/EHU, Apartado 644, 48080 Bilbao, Spain}

\author{Mateo Casariego}
\altaffiliation{mateo.casariego@ehu.eus}
\affiliation{Department of Physical Chemistry, University of the Basque Country UPV/EHU, Apartado 644, 48080 Bilbao, Spain}
\affiliation{Basque Center for Applied Mathematics (BCAM), Alameda de Mazarredo 14, 48009 Bilbao, Spain}
\affiliation{EHU Quantum Center, University of the Basque Country UPV/EHU, Bilbao, Spain}

\author{Gabriel Molina-Terriza}
\affiliation{Centro de Física de Materiales (CFM-{MPC}), CSIC-UPV/EHU, Paseo Manuel de Lardizabal 5, 20018 Donostia-San Sebastián, Spain}
\affiliation{Donostia International Physics Center (DIPC), Paseo Manuel de Lardizabal 4, 20018 Donostia-San Sebastián, Spain}
\affiliation{IKERBASQUE, Basque Foundation for Science, María Díaz de Haro 3, 48013 Bilbao, Spain.}

\author{Íñigo Luis Egusquiza} 
\affiliation{Department of Physics, University of the Basque Country UPV/EHU, Apartado 644, 48080 Bilbao, Spain}
\affiliation{EHU Quantum Center, University of the Basque Country UPV/EHU, Bilbao, Spain}

\author{Mikel Sanz} 
\affiliation{Department of Physical Chemistry, University of the Basque Country UPV/EHU, Apartado 644, 48080 Bilbao, Spain}
\affiliation{IKERBASQUE, Basque Foundation for Science, Plaza Euskadi 5, 48009 Bilbao, Spain}
\affiliation{Basque Center for Applied Mathematics (BCAM), Alameda de Mazarredo 14, 48009 Bilbao, Spain}
\affiliation{EHU Quantum Center, University of the Basque Country UPV/EHU, Bilbao, Spain}

\begin{abstract}
A rotating surface can induce a frequency shift in incident light by changing its angular momentum, a phenomenon known as the rotational Doppler effect. This effect provides a means to estimate the angular velocity of the rotating surface. In this work, we develop a continuous-variable quantum protocol for estimating the angular velocity of a rotating surface via the rotational Doppler effect. Our approach exploits squeezed and displaced Laguerre-Gaussian modes as quantum resources, which interact with a rotating metallic disc with surface roughness. The frequency shift induced by the rotational Doppler effect is then measured using a homodyne detection scheme. By analyzing the Fisher information, we demonstrate that the proposed squeezing-enhanced protocol achieves Heisenberg scaling in the ideal noiseless regime. Furthermore, we investigate the influence of noise and consider different surface models to assess their impact on the protocol's performance. While Heisenberg scaling is degraded in the presence of noise, we show that optimizing the energy allocation ratio between displacement and squeezing of the probe ensures that the quantum strategy consistently outperforms its classical counterpart.
\end{abstract}

\maketitle
\section{Introduction}\label{Sec1}

Most phenomena in light-matter interactions, including radiation pressure or the photoelectric effect, can be accounted for with light's linear momentum. But others, like the torque exerted by light on a suspended birefringent plate \cite{Beth1936}, need to incorporate a rotational contribution.
Angular momentum in light splits into spin and orbital contributions (SAM and OAM, respectively). Loosely speaking, SAM is present when the light field is circularly polarized, while OAM occurs when the field carries a non-trivial azimuthal dependence. The simplest of such dependences is a phase in the transverse plane that goes like $\exp(i l \phi)$ with $l$ an integer and $\phi$ the azimuthal angle. These `helically phased' beams carry an average Poynting vector that is tilted with respect to the beam axis, with a local skew angle on the order of  $ l/k r$ in the paraxial approximation, for a radial distance $r$, with $k = 2 \pi /\lambda$ and $\lambda$ the wavelength \cite{Leach:06}. A prominent modal basis for such beams is given by the Laguerre-Gaussian (LG) modes \cite{allen1992orbital}, whose first experimental demonstration dates back to 1994 \cite{beijersbergen1994helical}.

The (special) relativistic linear Doppler effect (RLDE) \cite{Einstein1905-OnTheElectrodynamics} is a shift in the observed frequency of the electromagnetic field that occurs when a light source and a detector move relative to each other along a line in flat spacetime. A perfect moving mirror reflects the light with a frequency shift proportional to its velocity relative to an observer. Motivated by this, it is natural to ask whether a similar effect can be observed when the relative motion is rotational, and whether this effect involves SAM, OAM, or both. 

However, the `rotational Doppler effect' (RDE) is not a relativistic effect --although, as some authors claim, both RLDE and RDE share a similar physical origin \cite{Fang_2017}, when rephrased in terms of conserved quantities--, but rather a consequence of light changing its angular momentum (OAM and/or SAM) upon interacting with matter that is rotating and/or spinning \footnote{The frequency shift can also be obtained via representing SAM (or a two-dimensional subspace of OAM) in a Poincaré sphere and invoking a Berry phase accumulation over a closed path (see A. M. Yao and M. J. Padgett, Advances in Photonics \textbf{3}, 161 (2011)). A generalization of this geometric argument to the full OAM space may be possible using Jones matrices instead (see D. Lopez-Mago et al Opt. Lett. \textbf{42}, 2667 (2017)).}. The RDE requires that the light interacts with structured matter: a perfect conducting mirror with rotational symmetry does not produce an observable shift in frequency, since its rotation is indistinguishable from the static case. In Ref.~\cite{lavery2013detection} the effect was demonstrated by observing a frequency shift proportional to $(l+\sigma)\,\Omega$ in the light scattered by a spinning object with angular velocity $\Omega$, where $l$ is the OAM, $\sigma= \pm 1$ for right and left-handed polarization (SAM). Importantly, the frequency shift is still present for parallel spinning and observation axes, {{due to macroscopic roughness of the material, i.e. in comparison with the wavelength of the incident light.}} 
More recently, other experiments were able to isolate the rotational contribution to the Doppler shift by combining two illuminating beams with opposite OAM \cite{liu2019experimental} {{, and even to remotely sense an object's rotation with the help of OAM-entangled pairs of photons \cite{Zhang_2019}.}} Recently, the RDE has found applications in velocity estimation~\cite{wan2025compact}, and it is expected to find use in gyroscope physics \cite{PhysRevLett.129.113901}.

In this article, we ask whether resorting to quantum properties of the illuminating beam can enhance the estimation of the rotation velocity of a classical object. We start by formally deriving the RDE as a frequency shift in Laguerre-Gaussian modes under the paraxial approximation, filling a gap in the classical literature. This allows us to rigorously quantize the theory, leading to a Bogoliubov transformation for the quantum field operators. We propose a protocol where these modes reflect off the rotating object (for which we propose two distinct models), and we retrieve the rotation velocity via homodyne detection. Two strategies are compared: with and without squeezing in the illuminating beam. For the first, which we call quantum, we use single-mode squeezed states, while in the classical strategy we consider coherent states. The performances are quantified by their respective Fisher information under the constraint of homodyne detection. We find that the quantum protocol achieves a quadratic improvement in the noiseless scenario (Heisenberg scaling), and we design a strategy where the energy allocation ratio between squeezing and displacement is optimized so that in the presence of noise the quantum protocol still outperforms its classical counterpart. This work could be relevant for experiments involving trapped microparticles with vortex light \cite{GomezViloria2024_OnAxisOpticalTrapping}, where a precise estimation of the rotation of the particle with few photons can be crucial to isolate other resonances and undesirable effects \cite{10.1063/5.0255437}.

The article is organized as follows: In Section~\ref{sec:Sec2}, by employing the classical electromagnetic wave reflection off a rotating surface under the paraxial approximation, we formally derive the RDE as a frequency shift in Laguerre-Gaussian modes; we then move to a quantum framework, and encode the RDE into an effective Bogoliubov transformation,  finding the input-output Hilbert space theory upon which we will base our quantum metrology strategy. This is discussed in Section~\ref{sec:3-EstimatingRotation}, where we describe a continuous variable multimode protocol with homodyne detection for the estimation of the rotation frequency, and find general formulae for the Fisher information. Results for two surface models are given in Section~\ref{sec:Results}: First, for a theoretical metasurface that induces a definite change in the orbital angular momentum, we find a Heisenberg scaling of the Fisher information when resorting to an input squeezed vacuum state in the noiseless regime; and second, for a reflective surface with small random defects, where we find a quantum advantage provided that we can engineer the energy allocation between displacement and squeezing. Finally, we conclude in Section~\ref{sec:conclusions}.

\section{Reflection by a rotating rough surface}\label{sec:Sec2}
\subsection{Classical rotational Doppler effect}
\begin{figure}[h]
    \centering
\includegraphics[width=0.48\textwidth]{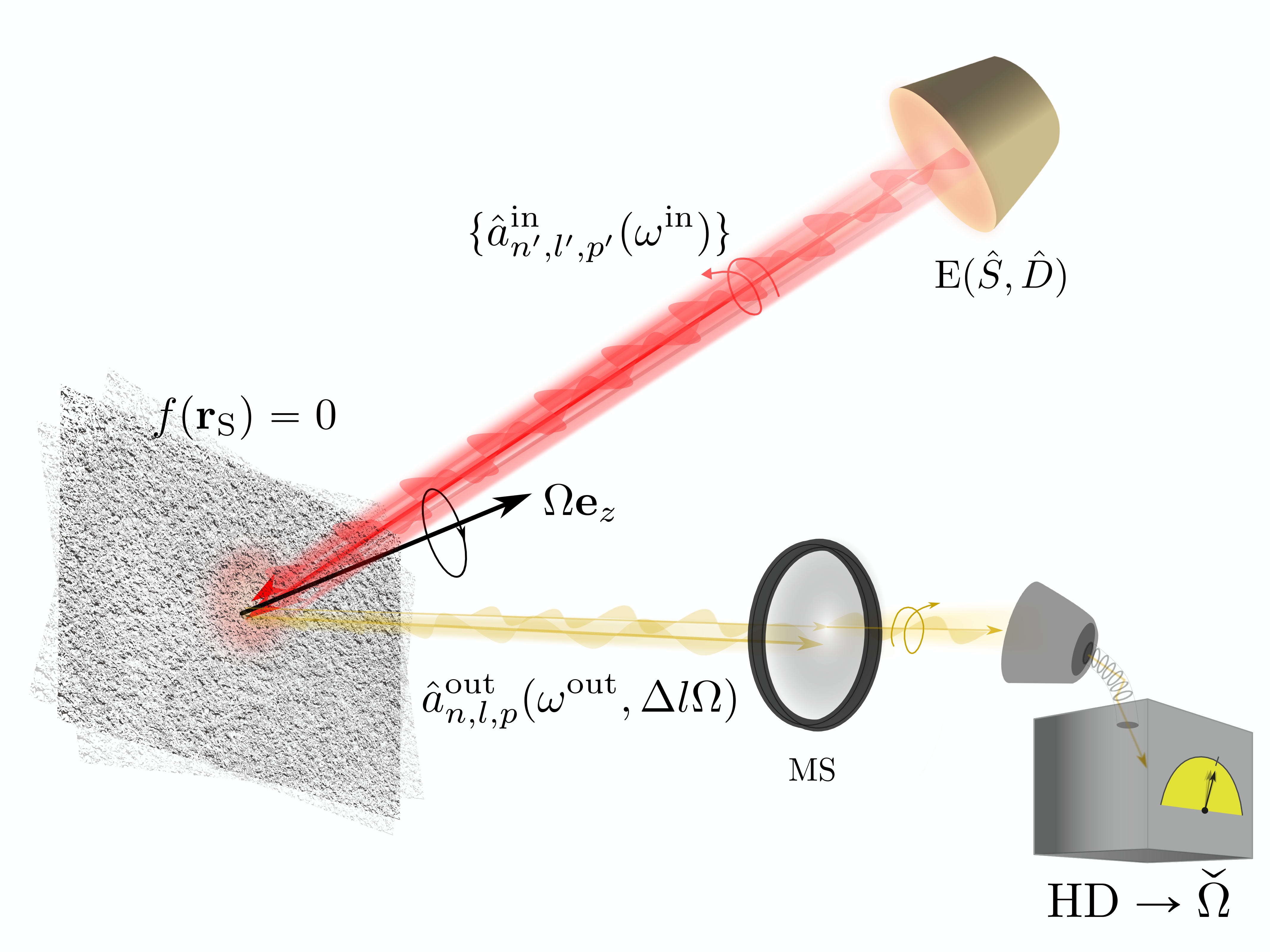}
    \caption{ Schematic recreation of our proposed squeezing-enhanced protocol for the estimation of the rotational Doppler effect. An emitter E can implement single-mode squeezing $\hat{S}$ and/or displacement $\hat{D}$ on some modes, producing a monochromatic propagating beam with field operators $\hat{a}^\text{in}_{n^\prime, l^\prime, p^\prime }(\omega)$, with the discrete subindices associated to a Laguerre-Gauss (LG) basis and where specifically $l$ corresponds to the quantized orbital angular momentum. The beam elastically interacts with a rough, flat surface characterised by some points such that  $f(\vec{r}_\text{S})=0$ with the function $f$ depending on the type of surface, which rotates with angular velocity $\Omega $ along the $z$ axis. For graphical convenience the propagation  and the rotation axes are not shown parallel, which is the simplified version studied in the main text. The outgoing, or scattered modes will be shifted in frequency through the change in OAM: $\Delta l \Omega$. Although the state after the interaction consists of several populated modes, only one of them is measured via a mode selector (MS). For simplicity, the measurement is homodyne detection (HD), whose outcomes are classically processed and an estimator $\widecheck{\Omega} $ for the angular velocity $\Omega$ is obtained.} 
    \label{fig:scheme}
\end{figure}

First we study the reflection produced by a non-rotating flat surface with some roughness (see Fig.~\ref{fig:scheme}) under paraxial incidence. 

We model the object as a mirror-like surface characterized by the points $\vec{r}_\text{S}$ that satisfy $f(\vec{r}_\text{S})=0$. Furthermore, we will eventually consider ensembles of surfaces, with the definition of the ensemble corresponding to the  roughness  of the surface. We assume that the average of the normal (unit) vectors is not zero, and choose that direction as (minus) the $z$ axis. This provides us with a definition of  perpendicular incidence, along that average normal, under which  the reflected wave's intensity exhibits an angular distribution, which arises from the electromagnetic field being a superposition of waves with different wave vectors. 

The Kirchhoff approximation \cite{jackson2021classical}, also known as the tangent plane approximation, is one of the most widely applied methods for analyzing scattering from surfaces. This approximation  assumes that the contribution to reflection at each point equals the one generated there  by an ideal, infinite flat surface tangent to the material's boundary. The approximation is valid if the radii of curvature at every point in the surface are large with respect to the incident wavelength.  

{{A rough surface model can be  compatible with the Kirchhoff approximation and/or with the weak scattering approximation. In what concerns the Kirchhoff approximation, it is required that the typical amplitude of a roughness fluctuation, $h$, and the typical distance between maxima $L$ combine so that $L^2/h$ be much larger that the wavelength of the impinging light.}}
In this context of the rough surface model, if \ILE{indeed} the Kirchhoff  approximation or the weak scattering approximation holds, the angular distribution of the reflected intensity broadens as the surface correlation length $L$ (interpreted as the typical distance between roughness defects) decreases. Conversely, for $L$ much larger than the beam wavelength, the intensity becomes predominantly non-zero in a single direction of space \cite{tsang2000scattering}, which entails that the reflected beam falls in the paraxial approximation, provided that it is illuminated by a beam also described by the paraxial approximation. That is, we illuminate the surface perpendicularly  with  Laguerre-Gaussian (LG) beams (see Appendix~\ref{sec:app} for details), with increasing $z$ corresponding to the propagation direction. For  a single LG mode, characterized by  $l$ and $p$, the incident electric field reads
\begin{align}
    \vec{E}_{l,p}^\text{in}(\vec{r},t)&=\bm{\epsilon}u^{in}_{l, p }(\vec{r})e^{-i(-k_z z+\omega t)}+ \text{c.c.},
\end{align}
 where $\bm{\epsilon}$ is a constant and homogeneous polarization unit vector field. In most of what follows we shall omit the coordinate argument of the spatial functions $u_{l,p}$. In the following, we assume that the electric field scattered by the surface remains within the paraxial approximation, by having curvatures and correlation lengths much larger than the wavelengths involved in the scattering. It follows that the  radiation scattered from incident paraxial waves   can also be written as a superposition of LG modes,
\begin{align}
\vec{E}^\text{out}(\vec{r})=&\bm{\epsilon}\sum_{l^\prime,p^\prime} c_{l^\prime,p^\prime}u_{l^\prime,p^\prime} e^{-i(k_z z+\omega t)}+c.c.\nonumber \\
+&\bm{\epsilon}^\perp \sum_{l^\prime,p^\prime} c_{l^\prime,p^\prime}^\perp u_{l^\prime,p^\prime} e^{-i(k_z z+\omega t)}+c.c.,
\end{align}
where  we note that the reflected beam propagates towards decreasing $z$. In its most general form the reflected beam could have different polarizations with amplitudes $c_{l^\prime,p^\prime} $ and $c_{l^\prime,p^\prime}^\perp $.
Assuming a perfect conductor with its surface characterized by $f(\vec{r}_\text{S})=0$ , the boundary condition for the electric field on the surface reads 
\begin{align}
    \vec{E}^\text{in}_\parallel(\vec{r}_\text{S})&+\vec{E}^\text{out}_{\parallel}(\vec{r}_\text{S})=0.
     \label{eq:boundaryconditions}
\end{align}
As shown in Appendix~\ref{app:polareq}, this implies for the case at hand, with a single $\lbrace l, p, \omega \rbrace$ LG mode in perpendicular incidence on a paraxiality preserving rough surface, that the reflected beam has the same polarization as the input beam, and Eq.~\eqref{eq:boundaryconditions} becomes the scalar equation
\begin{align}\label{eq:classicalIn-Out}
   u_{l,  p } e^{-i(-k_z z+\omega t) }&+\sum_{l^\prime,p^\prime} c_{l^\prime,p^\prime;l,  p }(\omega) u_{l^\prime,p^\prime}e^{-i(k_z z+\omega t)}=0
\end{align}
for the points such that $f(\vec{r}_\text{S})=0$.
The coefficients $c_{l,p;l^\prime, p^\prime }(\omega)$ are thus determined by the surface's properties. In principle, they are computable by observing that here $z$, on the surface, is a function of the transversal coordinates $x$ and $y$. Therefore, we have a functional equation that can be converted into a numerable system of algebraic equations by expanding in any numerable basis of functions on $\mathbb{R}^2$. In practice, it is advisable to use the LG basis itself, $\left\{u_{l,p}(x,y,\zeta)\right\}$, with $\zeta$ the average $z$ of the surface, and centered at the impact of the incident beam axis on the the surface. The coefficients $c_{l,p;l^\prime, p^\prime }(\omega)$ could be also computed by measuring the complex wavefunction of the scattered field. In \cite{PhysRevLett.115.193602} a simple method for complex wavefunction estimation is presented. We note that for surfaces that are invariant under rotations, \textit{i.e.} $ f(z,\phi,r)=f(z,\phi+\theta,r)$ for any $\theta$, the reflected beam would have the same OAM as the incident one, thus the RDE will not be observed. This leads to the necessity of incorporating {{either structured surfaces breaking the rotational symmetry (see Appendix~\ref{Roughmodel}), or roughness to the same effect, in order for RDE to be realised}}.

Once these coefficients are determined, we have an expression for both the incident and the scattered field away from the scattering surface. We now go to a rotating reference frame, with the rotation axis coinciding with the axis of the incident beam and the mean normal direction of the scattering surface. Let  $R(\Omega t)\in \mathsf{SO}(3)$ be the coordinate transformation  that relates the points in the rotating and inertial reference frames: $\vec{r}^\prime=R(\Omega t)\vec{r}$. In this reference frame the ``in'' field is
\begin{align*}
      &\tilde{\vec{E}}^\text{ in}_{l,p}=\bm{\epsilon}\:u_{l, p }(\vec{r^\prime})e^{-i(-k_z z+(\omega-l \Omega)t)} +c.c.,
\end{align*}
and the ``out'' field is
\begin{align*}
     &\tilde{\vec{E}}_{l,p}^\text{out}=\bm{\epsilon}\sum_{l^\prime,p^\prime} c_{l^\prime,p^\prime;l , p}(\omega)u_{l^\prime,p^\prime}(\vec{r}^\prime)e^{-i(k_{z} z+(\omega-l^\prime\Omega)t)}+c.c.
\end{align*}
where we used the fact that rotations in the $x-y$  plane keep the vector $\vec{e}_z$  invariant, and that  $u_{l^\prime,  p^\prime }(R(\Omega t)^{-1}\vec{r})=u_{l^\prime,   p^\prime }(\vec{r})e^{il^\prime \Omega t} $. 

Our objective now is to describe the physics in an inertial frame in which the scattering surface is rotating around the same axis, and the incident beam is a monochromatic Laguerre-Gaussian one of frequency $\omega^\prime$.  By inspection of the expressions above, it seems (with a subtlety), that this is achieved by defining $\omega^\prime=\omega-l^\prime\Omega$. The subtlety is that $k_z$ was understood earlier as  being $\omega/c$, and the incoming mode does not satisfy such a condition. However, as long as $\Omega/\omega$ is very small, in the hierarchy $\Omega/\omega\ll 1/k_zw_0\ll 1$, where $w_0$ is the waist radius of the paraxial beam, the paraxial approximation is valid (for further details, see Appendix~\ref{sec:app}). 

If we rename the frequencies $\omega-l^\prime \Omega\to \omega$ and the wave number $k_z'\to k_z$ the ``in'' and ``out'' fields seen by a reference frame where the surface is rotating are
\begin{align}
   &\tilde{\vec{E}}^\text{in}_{l,p}=\bm{\epsilon}\: u_{l , p }(\omega)e^{-i(-k_zz+\omega t)}+c.c.\label{eq:Ein}\\
     &\tilde{\vec{E}}_{l,p}^\text{out}= \bm{\epsilon}\sum_{l^\prime,p^\prime} c_{l^\prime,p^\prime;l , p }u_{l^\prime,p^\prime}e^{-i(k_{z}z+(\omega-\Delta l^\prime\Omega ) t)}+c.c.,
     \label{eq:Eout}
\end{align}
where $\Delta l=l^\prime -l$ is the change in the OAM. The shift $\Delta l \Omega$ experienced by the reflected field in Eq.~\eqref{eq:Eout} is the rotational Doppler shift \cite{emile2023rotational}.

\subsection{Rotational Doppler effect as a Bogoliubov transformation}
We have so far studied classical spatial beam modes and how the RDE appears and is expressed in those terms. Our interest, however, lies with quantum protocols for measuring the rotation's angular velocity, so it behoves us to pass to a quantum description. As is usual in beam quantum optics, given an orthogonal basis of beam states,  canonical creation and annihilation operators are associated with each such mode. Yet the spatial beam modes are not the most useful ones for our purposes, and temporal beam modes must be brought into play. As has been well established \cite{PhysRev.145.1041,raymer2020temporal} for general temporal (non-monochromatic) modes, in the beam context quantization of those temporal modes follows from that of spatial beam modes. As we have seen above that incident homogeneous and constant polarization is preserved, we will work with scalar fields for simplicity, and we choose to use Hermite-Gauss temporal modes with LG transverse profiles to represent our state. These modes arise naturally in light–matter interactions and form the basis used in many experimental platforms \cite{presutti2024highly,raymer2020temporal, Gonzalez_Raya_2025,PhysRevLett.99.243601}. One orthonormal family of such modes is given by
\begin{align*}
    \psi_{n,l,p}(t,r,\phi,z&;\tau_0, \omega_0, \theta_0, \sigma)=\\ &\Phi_n(t - z/c; \tau_0, \omega_0, \theta_0, \sigma)\cdot u_{l,p}(z,r,\phi)\,
\end{align*}
where a LG mode $u_{l,p}(z,r,\phi) $ is temporal modulated by an HG mode $\Phi_n(t - z/c; \tau_0, \omega_0, \theta_0, \sigma) $ see Appendix~\ref{sec:app}.
Each set of values
$ \{\omega,\tau_0,\theta_0,\sigma\}$ determines a full orthonormal basis and the subindices $\{n,l,p\}$ specify the different elements of that basis. 
The four parameters, $ \{\omega,\tau_0,\theta_0,\sigma\}$, correspond respectively to the central frequency of the wave packet, the center of the time profile of the wave, a global phase with respect to the lab reference frame, and the frequency width of the beam. 
Crucially, we use bases with different parameters to describe the input (incident) and output (scattered) beams, as the RDE can be understood in those terms. Thus, we label the central frequencies in these bases as $ \omega^\text{in}$ and $ \omega^\text{out}$, respectively and,
as a consequence of Eqs.~\eqref{eq:Ein} and ~\eqref{eq:Eout}, the relation between ``in'' and ``out'' annihilation operators can be written as 
\begin{align}\label{eq:InOutAnnihilationRelation}
\hat{a}^\text{out}_{n, l,  p }= \sum_{l^\prime,p^\prime,n^\prime} c_{l^\prime,p^\prime;l, p } K_{n^\prime;n }(\beta_{l^\prime,l })\hat{a}^\text{in}_{n^\prime,l^\prime,p^\prime}\\ \beta_{l,l^\prime}:=\frac{\omega^\text{in}-\Delta l\Omega-\omega^{\text{out}}}{\sigma}\nonumber,
\end{align}
where $ \hat{a}^\text{out}_{n, l,  p } $ is the annihilation operator of the ``out'' mode $ \psi_{n ,l , p }$ with commutation relation $ [\hat{a}^\text{out}_{n^\prime, l^\prime,  p^\prime },{\hat{a}^\text{out}_{n,l,p}}{}^\dagger]=\delta_{n,n^\prime}\delta_{l,l^\prime }\delta_{p, p^\prime }$. The function $K_{n^\prime;n}(\beta_{l^\prime,l})$ relates the ``in'' and ``out'' modes and can be understood as a consequence of changing from one base to another. The details of the derivation of Eq.~\eqref{eq:InOutAnnihilationRelation} can be found in Appendix~\ref{app:temporal}. 

\section{Estimating the rotation frequency}\label{sec:3-EstimatingRotation}
In this section we present our quantum strategy for the  estimation of the rotational velocity $\Omega$. A schematic view of the problem is depicted in Fig.~[\ref{fig:scheme}]. We shine on the rotating rough surface with an electromagnetic field. The reflected signal is received and a homodyne measurement is performed on a selected single mode, motivated by the experimental difficulty of performing global and non-Gaussian measurements. The experimental data is classically processed to find an estimator $\widecheck{\Omega}$ for the true value $\Omega$.

\subsection{Probe state and target mode}\label{strategy}
Denote the target measured mode by $\mathbf{m}:=\{n,l , p \}$.  This output mode is a linear combination of inputs, 
\begin{align}\label{eq:InputOutputFieldOperators}
&\hat{a}^\text{out}_\mathbf{m}=\sum_{\vec{i}}U_{\mathbf{m}\mathbf{i}}(\Omega)\:\hat{a}_\mathbf{i}^\text{in},
\end{align}
 with  coefficients $U_{\mathbf{m}\mathbf{i}}(\Omega)$. We assume as a starting point that a good prior of the value of the parameter is known \cite{gorecki2020,meyer2025quantum,navarro2025existence}. Thus we write $\Omega=\Omega_0+ \delta\Omega $ where $\Omega_0$ is the prior, and we want to estimate $\delta\Omega$ in the limit $\delta \Omega/\Omega_0 \ll 1$. Therefore, we can approximate the transformation to first order as $U(\Omega_0+\delta \Omega)\approx U(\Omega_0)G(\Omega_0)$ where $G(\Omega_0)= [\Id+U(\Omega_0)^{-1}\partial_\Omega U^\prime(\Omega_0) \delta \Omega ]$. To ease the notation, in the following we set $\tilde U=U(\Omega_0)$ and $\tilde G=G(\Omega_0)$ so,
\begin{align}
\hat{a}^\text{out}_{\vec{m}}=\sum_{\vec{j},\vec{i}} \tilde U_{\vec{m}\vec{j}}\tilde G_{\vec{j}\vec{i}}\hat{a}^\text{in}_{\vec{i}}.
\end{align}
We note that the approximated transformation is unitary up to first order in the parameter, meaning that the computed Fisher information is exact \cite{sorelli2024gaussian}. The explicit form of the matrices $\tilde G$ and $\tilde U$ can be found in Appendix~\ref{app:temporal}. The matrix $\tilde U$ determines the transformation around the estimation point $\Omega_0$ and it is necessary to design an optimal probe state. Note that when $\Omega\to \Omega_0$, that is to say, when $\delta \Omega\to0 $, then $\tilde G \to \Id$. Hence, we have that
\begin{align}
\hat{a}^\text{out}_\mathbf{m}=\sum_{\vec{i}}\tilde U_{\mathbf{m}\mathbf{i}}\hat{a}_\mathbf{i}^\text{in},\hspace{1cm} \text{for} \quad \Omega \to\Omega_0.
\end{align}
Now we assume that there is a countable set of values of $\mathbf{i}$ that satisfy $U_{\mathbf{m}\mathbf{i}}(\Omega)\neq0 $;  we call this set of modes $I$. Then,  if we restrict ourselves to displacement and single-mode squeezing, the most general quantum input or probe state is:
\begin{align}
|\Psi_\text{Q}\rangle^\text{in}=\bigotimes_{\mathbf{i}\in I}\hat{D}_{\mathbf{i}}(\alpha_\mathbf{i})\hat{S}_{\mathbf{i}}(\xi_\mathbf{i}) |0\rangle,
\label{eq:qstate}
\end{align}
that belongs to Hilbert space $\hilbert = \bigotimes_{n, l, p} \hilbert_{n,l,p}$, and where $\hat{D}_{\mathbf{i}}(\alpha_\mathbf{i})$ and $\hat{S}_{\mathbf{i}}(\xi_\mathbf{i}) $ are the displacement and  the squeezing operators, respectively (see Appendix~\ref{CV} for details).
The input state encompasses all modes that contribute to the measured mode $\mathbf{m}$, ensuring the generality of the approach.

Next, we take advantage of the symplectic formalism for Gaussian states in order to ease the characterization of this family of input states.

\subsection{Mean and covariance matrix }
Gaussian states are quantum continuous variable states \cite{serafini2023quantum,weedbrook2012gaussian}  characterized by their mean vector $\mathbf{r} \equiv \langle \hat{\mathbf{r}} \rangle$ and covariance matrix $\left[\Sigma\right]_{\vec{i} \vec{j}} = \frac{1}{2}\langle\hat{\mathbf{r}}_\vec{i} \hat{\mathbf{r}}^\trans_\vec{j} + \hat{\mathbf{r}}_\vec{j} \hat{\mathbf{r}}^\trans_\vec{i} \rangle - \langle \hat{\mathbf{r}}_\vec{i} \rangle \langle \hat{\mathbf{r}}_\vec{j}^\trans \rangle$. We use $ \hat{\mathbf{r}}= (\hat{\mathbf{r}}_\vec{1}, \ldots, \hat{\mathbf{r}}_\vec{d})^\trans=:(\hat{x}_\mathbf{1},\hat{p}_\mathbf{1},\ldots,\hat{x}_\mathbf{d},\hat{p}_\mathbf{d})^\trans$, which defines the `real basis'. See Appendix~\ref{CV} for further details and conventions on quantum continuous variables.

The considered probe state, Eq.~\eqref{eq:qstate}, is Gaussian with displacement 
$\mathbf{r} =(x_\mathbf{1},p_\mathbf{1}
\dots, x_\mathbf{d}, p_\mathbf{d})$ 
where $x_\mathbf{i}=2\Re[\alpha_\mathbf{i}]$, $p_\mathbf{i}=2\Im[\alpha_\mathbf{i}]$ and since it is a product state its  covariance matrix is block-diagonal, $\Sigma = \bigoplus_\mathbf{i}^d \Sigma_{\mathbf{i}} $, where for simplicity we have defined the $2\times 2$ matrix $\Sigma_{\mathbf{i}}\equiv \left[\Sigma\right]_{\mathbf{ii}} $. Given that $ \xi_{\mathbf{i}}=s_{\mathbf{i}}e^{i\theta_{\mathbf{i}}} $ for $\vec{i} \in I$ and $s_\vec{i}=|\xi_{\mathbf{i}} |$ the matrices $\Sigma_{\mathbf{i}} $ can be written:
\begin{align*}
& \Sigma_{\mathbf{i}}  =\begin{pmatrix}
\Sigma_{\mathbf{i}}^{xx} & \Sigma_{\mathbf{i}}^{xp}\\
\Sigma_{\mathbf{i}}^{xp} & \Sigma_{\mathbf{i}}^{pp}
\end{pmatrix}=\\
&\begin{pmatrix}
\cosh(2s_{\mathbf{i}})-\cos\theta_{\mathbf{i}}\,\sinh(2s_{\mathbf{i}}) & \sin\theta_{\mathbf{i}}\,\sinh(2s_{\mathbf{i}}) \\
\sin\theta_{\mathbf{i}}\,\sinh(2s_{\mathbf{i}}) & \cosh(2s_{\mathbf{i}})+\cos\theta_{\mathbf{i}}\,\sinh(2s_{\mathbf{i}})
\end{pmatrix}.
\end{align*}
In this formalism, the transformation of Eq.~\eqref{eq:InputOutputFieldOperators} is captured by a transformation matrix $T$ defined as
\begin{align*}
T =
\begin{pmatrix}
\Re[U_{11}] & -\Im[U_{11}]  &  \dots& -\Im[U_{1n}] \\
\Im[U_{11}] & \Re[U_{11}]  &  \dots& \Re[U_{1n}]\\
 \vdots &\vdots& \ddots& \vdots  \\
 \Im[U_{n1}] & \Re[U_{n1}] & \dots& \Re[U_{nn}] \\
\end{pmatrix}.
\end{align*}
Covariance and mean vector of the output state transform with $T$ via
\begin{align}
{\Sigma}^\text{out}&=T{\Sigma}T^\trans\\
\mathbf{r}^\text{out} &=T\mathbf{r}.
\end{align}
As we are focusing on one out-mode, the mode $\mathbf{m}$, we write the displacement vector and covariance for this mode,
\begin{align}
\mathbf{r}_{\mathbf{m}} =
\begin{pmatrix}
x_{\mathbf{m}} \\
p_{\mathbf{m}}
\end{pmatrix}= \sum_{\vec{i}\in I}\begin{pmatrix}
\Re[U_{\mathbf{m}\mathbf{i}} ]x_\mathbf{i}-\Im[U_{\mathbf{m}\mathbf{i}} ]p_\mathbf{i} \\
\Im[U_{\mathbf{m}\mathbf{i}} ]x_\mathbf{i}+\Re[U_{\mathbf{m}\mathbf{i}} ]p_\mathbf{i}
\end{pmatrix}.
\end{align}

As we explain below in Section\ref{HOmodyne}, we will measure a single quadrature. It is therefore interesting to focus all the information encoded in the displacement in that quadrature (see also Appendix~\ref{sec:appD} for more details). On the other hand, the covariance matrix of the relevant mode is
\begin{widetext}
\begin{align}
\Sigma_{\mathbf{m}} (\Omega)
=\sum_{\vec{i}\in I}\begin{pmatrix}
R_{\mathbf{m}\mathbf{i}}^2 \Sigma^{xx}_{\mathbf{i}}+I_{\mathbf{m}\mathbf{\mathbf{i}}}^2\Sigma^{pp}_{\mathbf{i}}-2R_{\mathbf{m}\mathbf{i}}I_{\mathbf{m}\mathbf{i}}\Sigma^{xp}_{\mathbf{i}}  &  (R_{\mathbf{m}\mathbf{i}}^2-I_{\mathbf{m}\mathbf{i}}^2)\Sigma^{xp}_{\mathbf{i}}+R_{\mathbf{m}\mathbf{i}}I_{\mathbf{m}\mathbf{i}}(\Sigma_\mathbf{i}^{xx}-\Sigma_\mathbf{i}^{pp})   \\
(R_{\mathbf{m}\mathbf{i}}^2-I_{\mathbf{m}\mathbf{i}}^2)\Sigma^{xp}_{\mathbf{i}}+R_{\mathbf{m}\mathbf{i}}I_{\mathbf{m}\mathbf{i}}(\Sigma_\vec{i}^{xx}-\Sigma_\mathbf{i}^{pp}) & R_{\mathbf{m}\mathbf{i}}^2 \Sigma^{pp}_{\mathbf{i}}+I_{\mathbf{m}\mathbf{i}}^2\Sigma^{xx}_{\mathbf{i}}+2R_{\mathbf{m}\mathbf{i}}I_{\mathbf{m}\mathbf{i}}\Sigma^{xp}_{\mathbf{i}}
\end{pmatrix},
\end{align}
\end{widetext}
where we have used that the input is a product state and we have defined $R_{\mathbf{m}\mathbf{i}}=\Re[U_{\mathbf{m}\mathbf{i}}]$, $I_{\mathbf{m}\mathbf{i}}=\Im[U_{\mathbf{m}\mathbf{i}}]$, which are the objects that carry the $\Omega$ dependency.

A quantum advantage can be found if the measured state has a quadrature with fluctuations below the shot-noise \cite{Slussarenko2017Unconditional}. It can be seen that the variance of quadrature $\hat{x}_\mathbf{m}$ is a sum of variances of rotated quadratures $\hat{x}_\mathbf{i} $. If we set the squeezing phase in the optimal angle $\theta_\mathbf{i}=\arctan(\frac{-\Im[\tilde U_{\mathbf{m}\mathbf{i}}]}{\Re[\tilde U_{\mathbf{m}\mathbf{i}}]})$ the covariance matrix at the point $\Omega=\Omega_0$ is
\begin{align}
\Sigma_{\mathbf{m}}(\Omega_0)= \sum_{\vec{i}\in I}
\begin{pmatrix}
 |\tilde U_{\mathbf{m}\mathbf{i}}|^2 e^{-2s_\mathbf{i}}&  0  \\
0 & |\tilde U_{\mathbf{m}\mathbf{i}}|^2 e^{2s_\mathbf{i}}
\end{pmatrix},
\end{align}
as can be seen all the squeezing is concentrated in the $\hat{x}$ quadrature.\\
We model the impact of noise by mixing the mode to be measured in a beam splitter with reflectivity $\eta \in [0,1]$ with an environment. At room temperature, the environment is always in a   thermal state, with an average photon number that highly depends on the frequency: for instance, a $5 \text{GHz}$ field the average photon number at $T\sim 300K$  is about 1250 \cite{Casariego2022,gonzalez2024satellite}, but at $500 \text{THz}$ this number is just $2\times 10^{-35}$. Our work is intended to find applications in optical frequencies, so we will approximate the thermal state by a vacuum \cite{reichert2022}. The beam splitter induces the following transformation in the quadratures,
\begin{align*}
    \begin{pmatrix}
\hat{x}_\mathbf{m} \\
\hat{p}_\mathbf{m} \\
\hat{x}_E \\
\hat{p}_E
\end{pmatrix}
\rightarrow
\begin{pmatrix}
\sqrt{1 - \eta} & 0 & \sqrt{\eta} & 0 \\
0 & \sqrt{1 - \eta} & 0 & \sqrt{\eta} \\
-\sqrt{\eta} & 0 & \sqrt{1 - \eta} & 0 \\
0 & -\sqrt{\eta} & 0 & \sqrt{1 - \eta}
\end{pmatrix}
\begin{pmatrix}
\hat{x}_\mathbf{m} \\
\hat{p}_\mathbf{m} \\
\hat{x}_E \\
\hat{p}_E
\end{pmatrix}
\end{align*}
where the subindex $E$ is for environment. Then the covariance matrix and mean vector of mode $\mathbf{m}$ after the mixing will be,
\begin{align*}
\Sigma_{\mathbf{m}} 
\to(1-\eta)\Sigma_{\mathbf{m}} +\eta\begin{pmatrix}
1  &  0   \\
 0 & 1 
\end{pmatrix},
\end{align*}
\begin{align*}
\mathbf{r}_{\mathbf{m}} \to\sqrt{(1-\eta)}\,\mathbf{r}_{\mathbf{m}}.
\end{align*}

\subsection{Fisher information and measurement scheme}\label{HOmodyne}

The ultimate precision achieved (asymptotically) by any unbiased estimator $\widecheck{\Omega}$ in the independent, identically distributed (i.i.d.) setting \footnote{This is compatible with having a single copy of the object as long as there is no measurement back-action (\textit{i.e.}, assuming macroscopic classicality of the rotating object), a condition that allows to map a sequential probing to an i.i.d. situation where we can apply the usual parameter estimation methods. The situation where the object is small enough to either be affected by the probe or to have quantized rotational degrees of freedom itself is left for future work.} is bounded by the Cramér-Rao bound (CRB), that sets a limit to its variance: $\text{Var}\; \widecheck{\Omega}\geq 1/F$, where $F$ is the classical Fisher information (CFI), which quantifies the dependence of the probability distribution of the experimental data, $P_\vec{x}(\Omega)$, with the parameter that is being estimated: $F = \int_X \diff \vec{x} P_\vec{x}(\Omega) (\partial_\Omega \ln P_\vec{x}(\Omega))^2$ \cite{CasellaBerger, Sidhu_2020}.  In quantum parameter estimation we need to account for the fact that the probability distributions associated to a parameter $\Omega$ emerge from  Born's rule: for an outcome $\vec{x}$ with continuous spectrum $\vec{x}\in \mathrm{X}\subseteq \reals $, we have to replace $P_\vec{x}(\Omega)$ by a conditional probability density $p_\vec{x}(\Omega | \rho_\Omega, \mathcal{M})\diff \vec{x} = \Tr (\rho_\Omega \hat{M}_\vec{x})\diff \vec{x}$, where $\mathcal{M} = \lbrace \hat{M}_\vec{x}\diff \vec{x} \rbrace_\vec{x}$ is a positive operator-valued measure (POVM) with effects $\hat{M}_\vec{x}\diff \vec{x}$. Note that $\int \hat{M}_\vec{x}\diff \vec{x} = \Id $. The probability densities depend on the measured state $\rho_\Omega$, that carries a parametric dependence on $\Omega$. This is encoded in the initial probe state $\rho$ through some quantum channel: $\rho_\Omega = \Lambda_\Omega [\rho]$, that codifies the probe-object interaction. With this, the CFI becomes a functional of the probability densities $p_\vec{x}(\Omega | \rho_\Omega, \mathcal{M})\diff \vec{x}$. Importantly, it depends on the change of $p_\vec{x}$ with the parameter $\Omega$ through $(\partial_\Omega \ln p_\vec{x})^2$. This means that different initial states $\rho$ will have different CFI, which is interpreted as an optimization opportunity in quantum metrology. The task is to find initial states with a Fisher information as large as possible. Or, in metrology words: to find a probe state that is very sensitive to the quantum channel under consideration.

The quantum Cramér-Rao bound (QCRB) is obtained as an further optimization of the CRB over POVMs such that the CFI is maximal (and the variance associated to the estimator minimal). The QCRB reads $(\Delta \widecheck{\Omega})^2\geq 1/J$, where $J = \max_{\mathcal{M}} F $ is the quantum Fisher information (QFI) \cite{toth2014quantum,paris2009}.

Strategies whose input consists of a product of coherent states of the following type will be denoted `classical':
\begin{align}
|\Psi_\text{C}\rangle^\text{in}=\bigotimes_{\mathbf{i}\in I}\hat{D}_{\mathbf{i}}(\alpha_\mathbf{i})|0\rangle.
\label{15}
\end{align}
Note that this state is a particular case of Eq.~\eqref{eq:qstate} when all $s_\mathbf{i}=0$. Moreover, in the next sections we will make explicit the distinction between the displacement parameters associated to the quantum and classical strategies, since, while they may apply to the same mode, their magnitude will not be equal if we want to make a fair energetic comparison between the strategies. This input state will be used to benchmark other strategies that use squeezing as in  Eq.~\eqref{eq:qstate} and that will be called `quantum'. For a fair benchmark, quantum strategies will not use more resources than the classical. For the estimation of an angular velocity, the resources are proportional to: the energy of the probe state, its time width, and the change in the angular momentum induced by the rotating surface (see Appendix \ref{sec:appD}).

In what follows, however, we will not need POVM optimization: we will be working strictly with classical Fisher information of homodyne as the fixed measurement scheme, whose performance is optimal for all classical strategies, provided that we can tune the phases of the displacement correctly (a task that may be experimentally demanding). This is shown in Appendix~\ref{sec:appD} and can be understood as particular instance of a larger class of estimation tasks \cite{navarro2025super}.

Homodyne detection is the process where a quadrature of the electromagnetic field is measured.  Homodyne detection can be mathematically modeled as a projection onto the eigenbasis of the quadrature whose expected value we seek to infer. Taking the position quadrature $\hat{q}$, the POVM is 
$\{ |q\rangle \langle q| \diff q\}_{q \in \mathbb{R}}$ where $|q\rangle $ are the eigenstates of $\hat{q} $. For a pure state $\ket{\Psi}$, the probability of obtaining the outcome $q$ is given by $p(q)=|\langle q|\Psi\rangle |^2$. Since homodyne detection is a Gaussian measurement --meaning that it preserves the Gaussianity of the state--, the probability distribution $p(q)$ is Gaussian, with mean $\bar{q}=\mathbf{r}_{\mathbf{m}}^{x} $ and variance $\Sigma_q= \Sigma_{\mathbf{m}}^{xx}$ for mode $\vec{m}$ \cite{serafini2023quantum}. Note that no information is recovered from the momentum quadrature. For that reason it is interesting to focus the squeezing and the displacement in the measured quadrature.\\
The CFI for a Gaussian distribution is   \cite{cenni2022thermometry,van2004detection},
\begin{equation}
    F = \partial_{\Omega} \bar{q}  \Sigma_q^{-1}  \partial_{\Omega} \bar{q} + \frac{1}{2} \text{Tr} \left[ \Sigma_q^{-1}  \partial_{\Omega} \Sigma_q  \Sigma_q^{-1} \partial_{\Omega} \Sigma_q \right]. \label{CFisher}
\end{equation}

So far we have considered states with single-mode squeezing and displacement in an arbitrary number of modes. However, in the following sections we will study instances of the state in Eq.~\eqref{eq:qstate} with just two populated modes, making them more experimentally feasible.

\section{Results}\label{sec:Results}
In this section we consider two different surfaces and different input states. We present some regimes where a quantum advantage is found for simple states even in the presence of noise.
\subsection{Metasurface}
\begin{figure}[ht!]
    \centering
\includegraphics[width=0.52\textwidth]{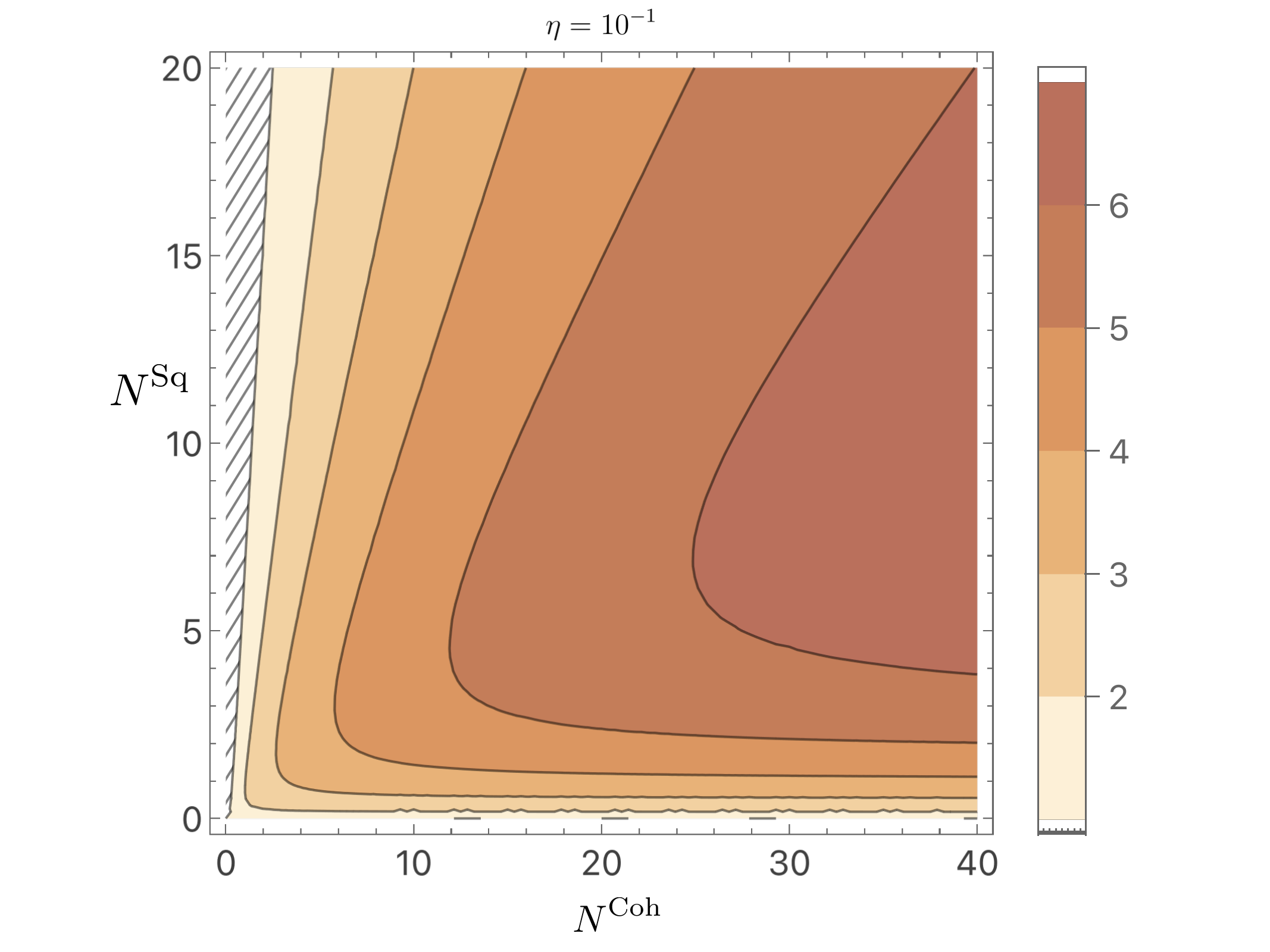}
    \caption{Contour plot of the ratio $R$ as a function of the squeezing and displacement photon numbers, $N^{\text{Sq}}$ and $N^{\text{Coh}}$, respectively. The ratio $R$ summarizes the quantum advantage of our squeezing-enhanced protocol for the rotational Doppler effect with a metasurface that induces a definite change in the incident OAM (see Eq.~\eqref{eq:RatioMeta}). We choose a realistic noise parameter $\eta=0.1$.  Note that for a given $N^{\text{Coh}}$ increasing the squeezing photon number $N^{\text{Sq}}$ could be detrimental for the advantage. The energy employed in the squeezing helps reducing the noise when measuring the displacement that is proportional to $N^{\text{Coh}}$. But if  $N^{\text{Coh}}$ is small, it is not optimal to use a lot of energy to reduce its noise.}\label{figN}
\end{figure}

Sometimes it is possible to engineer the body whose rotation we want to estimate, for example for gyroscope applications \cite{passaro2017gyroscope}. Recently there has been an enormous effort in designing and creating surfaces capable of manipulating the OAM of the light \cite{sroor2020high}. These surfaces are usually called `metasurfaces', for a review see Ref.~\cite{ahmed2022optical}.  Here we study the achievable precision when the surface is a metasurface that induces a change in the OAM of $\Delta l^*$ of the input beam,
\begin{align}
      c_{l^\prime,p^\prime;l,  p }=-\delta_{l^\prime,l +\Delta l^*} \delta_{p^\prime, p }.
\end{align}
We note that constructing a surface that exactly implements this transformation on the output field may be technologically challenging. However, it may be possible to design a surface that approximates this transformation for a restricted set of input modes ($l^\prime,p^\prime$).
For this surface model the mode transformation of Eq.~\eqref{eq:InOutAnnihilationRelation} and \eqref{eq:InputOutputFieldOperators} reads to first order in $\delta\Omega$ as follows:
\begin{align}
      \hat{a}^\text{out}_{n,l,  p }&=\nu_l\hat{a}^\text{in}_{n-1,l +\Delta l^*, p }+ (1+\mu_l)\hat{a}^\text{in}_{n,l +\Delta l^*, p }\nonumber\\
      &
    +\gamma_l\hat{a}^\text{in}_{n+ 1,l +\Delta l^*, p } \label{eq:20}
\end{align} 
where $  \nu_l= -i \frac{\sqrt{n}}{\sigma}\Delta l^* \delta\Omega$, $  
\mu_l =-i\tau\Delta l^* \delta\Omega $ and $  \gamma_l = -i \frac{\sqrt{n+1}}{\sigma}\Delta l^* \delta\Omega$. In words Eq.~\eqref{eq:20} states that the annihilation operator $\hat{a}^\text{out}_{n,l,p}$ is a linear combination of ``in'' annihilation operators with amplitudes that depend on the quantity to be estimated $\delta\Omega$. The derivation of Eq.~\eqref{eq:20} can be found in Appendix~\ref{app.Meta}.
To estimate $\delta\Omega$  we study the following input state:
\begin{align}
    |\Psi_\text{Q}\rangle^\text{in}=\hat{D}_{n+1,l-\Delta l^*,p}(\alpha_\text{Q})\hat{S}_{n,l-\Delta l^*,p}(\xi)|0\rangle.
\end{align}
The motivation for using such a state comes from experimental reasons: a vacuum squeezed state is more feasible than a squeezed displaced state. That is why displacement and squeezing are placed in different modes. In Appendix~\ref{app.Meta} it is shown that the Fisher information of the quantum strategy is
\begin{equation}\label{eq:quantumFI}
    F_\text{Q} = \frac{4(1-\eta) |\alpha_\text{Q}|^2 \frac{n+1}{\sigma^2}(\Delta l^*)^2  }{(1-\eta) e^{-2s}+\eta}.
\end{equation}
In the noiseless scenario $ \eta=0$ it can be shown that Eq.\eqref{eq:quantumFI} asymptotically reaches the Heisenberg scaling (see Appendix~\ref{app.Meta}). For completeness in Appendix~\ref{sec:appG} we compute the QFI and conclude that in general homodyne is not the optimal measurement.
Since here the set of modes $I$ consists of a single mode only, the optimal classical strategy for this transformation is provided by the input state
\begin{align}
    |\Psi_\text{C}\rangle^\text{in}=\hat{D}_{n+1,l +\Delta l^*, p }(\alpha_\text{C})|0\rangle,
\end{align}
which gives the following Fisher information (see Appendix~\ref{app.Meta}),
\begin{equation}\label{eq:classicalFI}
         F_\text{C}=4(1-\eta) |\alpha_\text{C}|^2 \frac{n+1}{\sigma^2}(\Delta l^*)^2 .
\end{equation}
Taking into account the relation between squeezing parameter and expected photon number, $$e^{-2s}=1+ 2N^\text{Sq}  - 2\sqrt{ N^\text{Sq} \left( N^\text{Sq} + 1 \right) },$$ and using $N^\text{Coh}=|\alpha_\text{Q}|^2  $ and $N=|\alpha_\text{C}|^2$ with $N =N^\text{Coh} + N^\text{Sq} $, we find that the ratio of Fisher information becomes:
\begin{align*}
\frac{F_\text{Q}}{F_\text{C}} =\frac{N^\text{Coh}}{N} \frac{1}{(1-\eta)\left( 1+ 2N^\text{Sq}   - 2\sqrt{ N^\text{Sq} \left( N^\text{Sq} + 1 \right) }\right)+\eta} 
\end{align*}
where for a fair comparison we have assumed 
that $\Delta l$ and $\sigma$ are the same in both expressions and $N $ is the number of photons of the classical strategy. In Fig.~\ref{figN}, we plot $F_\text{Q}/F_\text{C}$ for different $N^{\text{Coh}}$ and $N^{\text{Sq}} $. Sometimes, arbitrarily increasing the squeezing can lead to a reduction in the quantum advantage.

We define $R$ as maximum quantum advantage for a fixed number of photons $N$,
\begin{equation}\label{eq:RatioOptimization}
R=\max_{N^\text{Sq},N^\text{Coh}}   \frac{F_\text{Q}}{F_\text{C}}  \quad \textrm{s.t.} \quad N=N^\text{Sq}+N^\text{Coh},
\end{equation}
which quantifies the maximum possible advantage provided by single-mode squeezing and homodyne detection. The optimization is performed with respect to the energy allocation, quantified by a fixed $N = N^\text{Sq}+ N^\text{Coh}$. A direct calculation yields 
\begin{equation}\label{eq:RatioMeta}
    R= \begin{cases}
        \frac{1+2N\eta - \sqrt{1+4N(1-\eta)\eta}}{2N\eta^2} \quad &\text{for } 1 \geq \eta>0\\
        1+N \quad &\text{for } \eta=0
    \end{cases}
\end{equation}
constrained by a specific function $N^\text{Coh} = N^\text{Coh}(N, \eta)$ whose details can be found in the Appendix~\ref{app.Meta}. 

Note that $R\geq1$ since the quantum strategy reduces to the classical strategy for $N^\text{Sq}=0$, as expected.

\begin{figure}[h!]
    \centering
\includegraphics[width=0.48\textwidth]{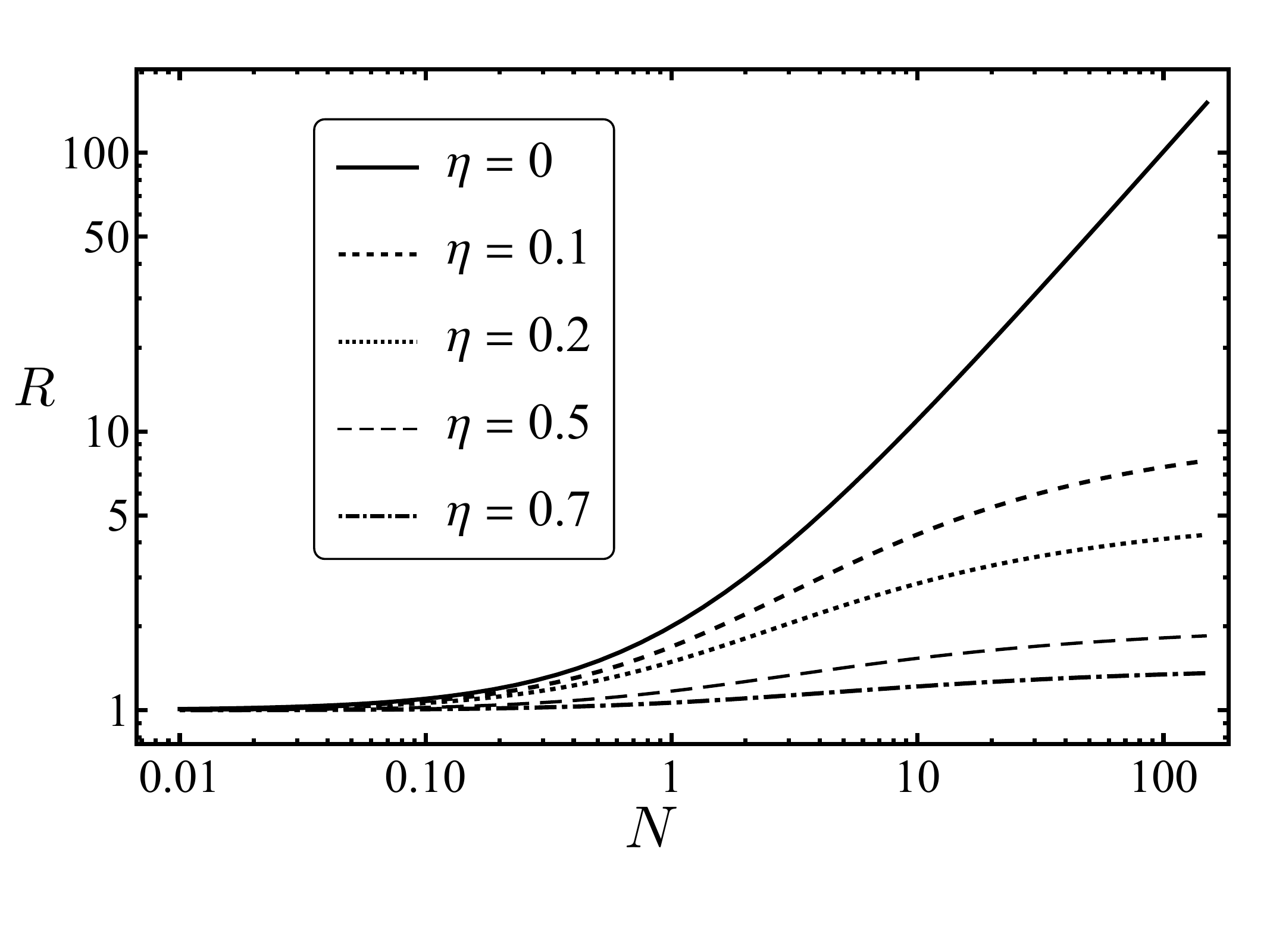}
    \caption{Ratio $R$ between the classical Fisher information  of our quantum strategy versus the optimal classical one, as a function of the total photon number $N$ --with optimal squeezing and displacement allocation--, and plotted for different noise levels $\eta$, including the noiseless case. $R$ increases fast for small photon number, showing a constant asymptotic behavior when $N$ is large. This maximum achievable value of $R$ for large $N$ decreases with $\eta$. However, as one should expect, this is not the case for the noiseless case, where $R$ shows the Heisenberg scaling described in the main text. 
    }
    \label{Metasurface N}
\end{figure}

\begin{figure}[h!]
    \centering
\includegraphics[width=0.48\textwidth]{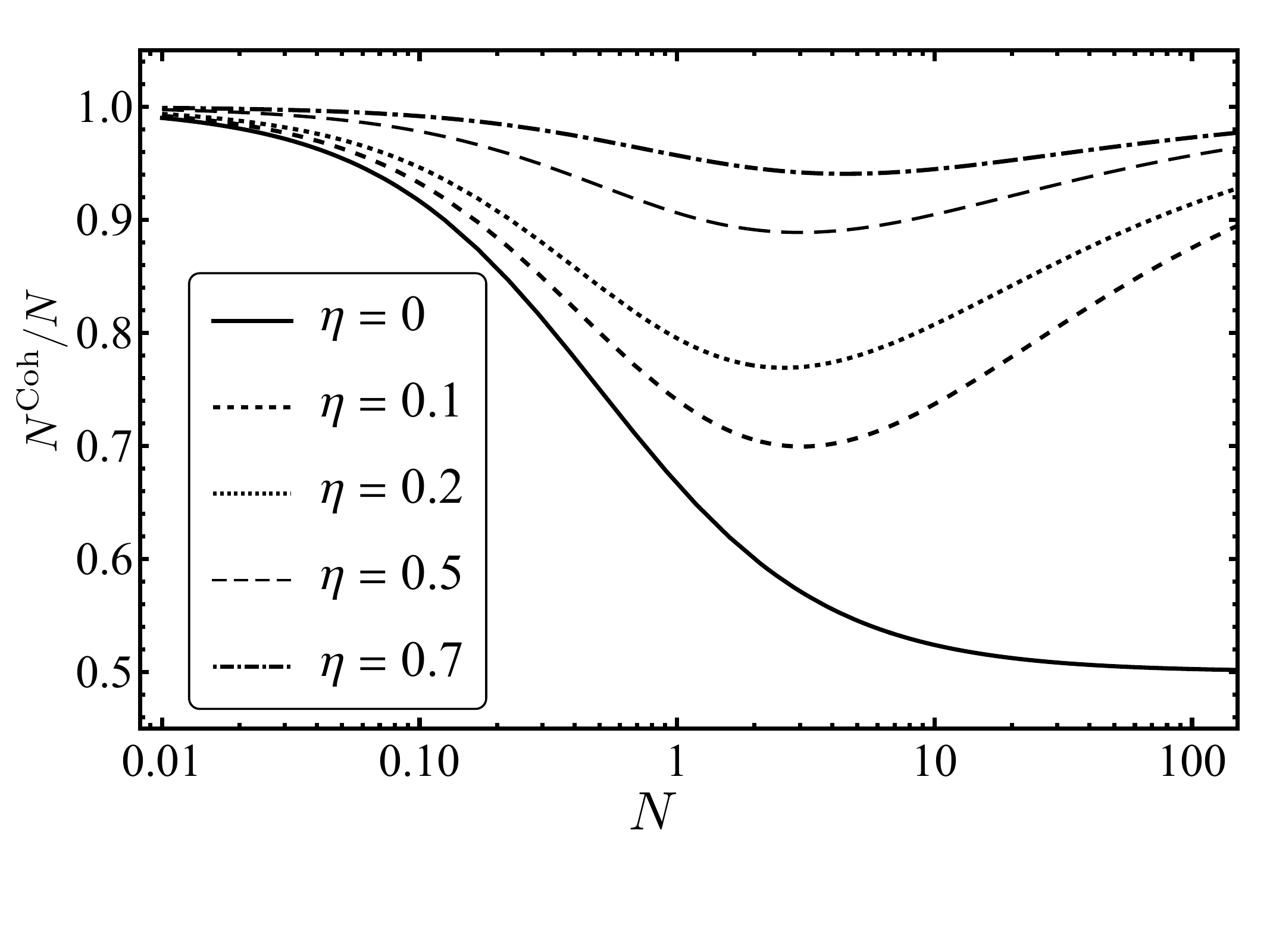}
    \caption{Optimal energy allocation, represented by the ratio $N^\text{Coh}/N$, as a function of the total number of photons $N$ for five different path transmittivities, $\eta$, including the ideal noiseless case. The optimal $N^\text{Coh}$ is determined from the optimization presented in Eq.~\eqref{eq:RatioOptimization}. We observe that for all plotted values of $N$ and $\eta$, the optimal probe state allocates more energy to displacement than to squeezing. This trend is particularly pronounced for higher values of $\eta$, indicating that the classical strategy --that is, without squeezing-- becomes nearly optimal in noisy environments. Interestingly, there seems to be a region around $N \approx 3$ for which squeezing allocation becomes crucial. The ideal noiseless scenario differs in the bright beam situation: for beams with $N \gtrsim 10$ the trend is that $N^\text{Coh}/N \rightarrow 0.5$, meaning that half of the resources should be put in squeezing and half in displacement.}
    \label{photonratio}
\end{figure}

We claim a `quantum advantage' in the estimation of $\Omega$ when this ratio is larger than one \cite{sanz2017quantum, casariego2020}. In Fig.~[\ref{Metasurface N}] we plot the ratio $R$ with respect to the total number of photons. As the total number of photons $N$ increases the quantum advantage increases reaching asymptotically a constant advantage that depends inversely on the noise $\eta$. As expected the noise is detrimental for any advantage. In the noiseless scenario $\eta=0$, since the Fisher information of the quantum strategy is proportional to $N^2$ the quantum advantage is proportional to $N$ and in principle it can be made arbitrary big. Nevertheless, the are two reasons that prevent us for having an unbounded advantage, first the noiseless scenario is a theoretical idealization not attainable in real life experiments. Second, highly populated squeezed states are difficult to generate, hence we plot for $N \leq 100$. In Ref.~\cite{vahlbruch2016detection} the authors detected a squeezed state with $15$dB of noise reduction with respect to the vacuum fluctuations, at an near-infrared wavelength of $1064$ nm. This state corresponds to a $N^\text{Sq}\approx7.4$ squeezed vacuum. In Fig.~\ref{photonratio} we plot how the photons are located in the optimal strategy. We note that always most of the photons are placed in the displacement. As already said this eases the preparation of the probe state. When the noise increases more photons should be placed in the displacement and less energy in the squeezing showing that in noisy regimes the optimal strategy is almost ``classical''.

\begin{figure}[b!]
    \centering
\includegraphics[width=0.48\textwidth]{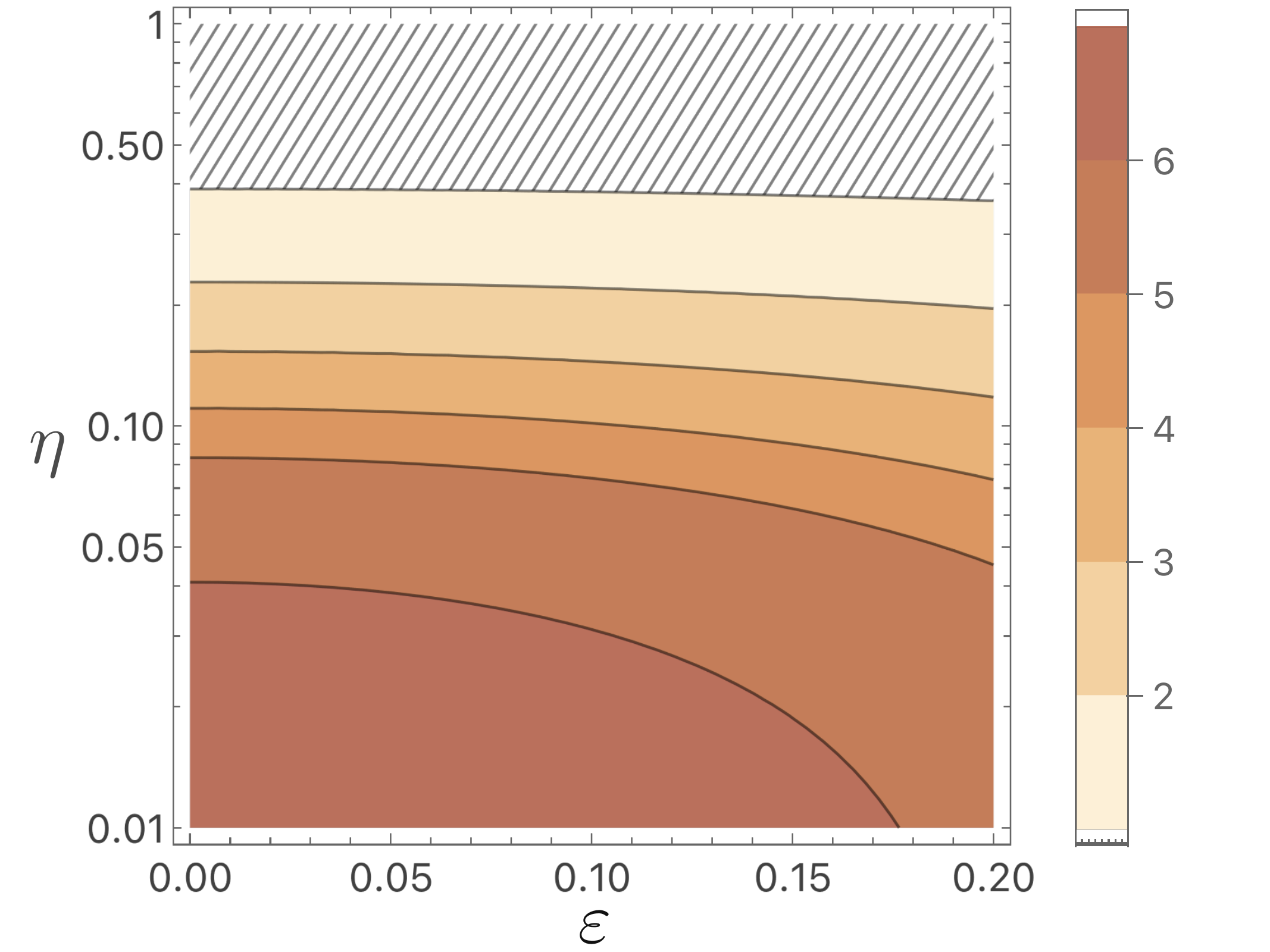}
    \caption{Gain maximization $R$, that is, the optimized ratio of Fisher information for the quantum vs the classical strategies (see Eq.~\eqref{ratiocomplex}) as a function of the transmittitivity $\eta$ (0 for noiseless, 1 for completely noisy) and the surface parameter $\varepsilon$ plotted for an expected total photon number $N=N^\text{Sq} + N^\text{Coh}=20 $. The parameter $\varepsilon$ is related to the energy of the scattered light that is diffused, and we assume $\varepsilon \in [0,0.2]$, more details about the model can be found in Appendix~\ref{Roughmodel}. The shaded region represents $R=1$, that is, no quantum advantage. As expected, $R$ grows inversely to both $\eta$ and $\varepsilon$.}\label{fig4}
\end{figure}

\subsection{Complex surface}
The previous model may not be adequate in certain situations. In particular, one may be interested in the angular velocity of an object whose surface exhibits a random roughness profile of small amplitude according to an appropriate metric. Such surfaces are commonly modeled as scattering most of the incident energy, $(1-\varepsilon^2)$, in a specular manner, while the remaining fraction, proportional to $\varepsilon^2$, is scattered diffusely as a consequence of the surface roughness. The roughness is typically modeled as a Gaussian random process, which leads to a Gaussian power spectral density of the scattered light \cite{goodman2015statistical}. Inspired by this we propose a Gaussian model for the scattering matrix,
\begin{align*}
     c_{l^\prime,p^\prime;l, p } =&\\ (-\sqrt{1-\varepsilon^2}&-i\frac{\varepsilon}{\mathcal{N}})\:\delta_{l^\prime,l }\delta_{p^\prime, p }+i\varepsilon \frac{ e^{-\frac{(l^\prime-l )^2}{4\sigma_l^2}} e^{-\frac{(p^\prime- p)^2}{4\sigma_p^2}}}{\mathcal{N}}e^{i\theta_{l, p;l^\prime ,p^\prime}},
\end{align*}
where $\mathcal{N}$ is a normalization factor whose details can be found in Appendix~\ref{Roughmodel}, the phase is antisymmetric $\theta_{l,p;l^\prime ,p^\prime} = -\theta_{l^\prime, p^\prime;l,p}$, the parameter $\varepsilon$ determines the partition of the energy of the reflected beam that is diffused and the parameter $\sigma_l$ determines how wide is the angular and radial distribution of the scattered light. This model is derived and further explained in Appendix~\ref{Roughmodel}.  The matrix $\tilde U_{\vec{i}\vec{j}}$ of the transformation is given by
\begin{align*}
\tilde U_{\vec{i}\vec{j}}&=\tilde U_{n^\prime,l^\prime,p^\prime;n,l,p}=K_{n^\prime;n}(\beta_{l^\prime,l })c_{l^\prime,p^\prime;l,  p }\\ \beta_{l,l^\prime}&=\frac{\omega^\text{in}-\Delta l\Omega-\omega^{\text{out}}}{\sigma}\,.
\end{align*}

Theoretically, the set $I$ has an infinite cardinality. Nevertheless it is not reasonable to suppose that the input state consist in an infinite number of squeezed modes. In the following we will study the performance of the input state,
\begin{align}
|\Psi_\text{Q}\rangle^\text{in}=\hat{D}_{n+1,l +\Delta l, p }(\alpha_\text{Q})\hat{S}_{n,l , p }(\xi)|0\rangle.
\end{align}
The Fisher information of this state is computed in Appendix~\ref{app.ComplexSurface} and reads:
\begin{equation}
    F_\text{Q} =\frac{1}{\mathcal{N}^2}\frac{[2\varepsilon e^{-\frac{(\Delta l)^2}{4\sigma_l^2}}  \Delta l\frac{n+1}{\sigma} ]^2|\alpha_\text{Q}|^2(1-\eta)}{(1-\eta)(1-\varepsilon^2)e^{-2s}+(1-\eta)\varepsilon^2+\eta}.
\end{equation}

We compare this Fisher information with the one given by the state
\begin{align}
|\Psi_\text{C}\rangle^\text{in}=\hat{D}_{n+1,l+\Delta l, p }(\alpha_\text{C})|0\rangle
\end{align}
that is,
\begin{equation}
    F_\text{C} = \frac{1}{\mathcal{N}^2}[2\varepsilon e^{-\frac{(\Delta l)^2}{4\sigma_l^2}}  \Delta l\frac{n+1}{\sigma} ]^2|\alpha_\text{C}|^2(1-\eta)
\end{equation}

We can analytically maximize the ratio of Eq.~\eqref{eq:RatioMeta} and find
\begin{widetext}
    \begin{align}
R= \begin{cases}
\frac{
1 - \sqrt{
1 + 4 N \, (1 - \varepsilon^2) \, (\varepsilon^2 (1 - \eta) + \eta) \, (1 - \eta)
} + 2 N \, (\varepsilon^2 (1 - \eta) + \eta)
}{
2 N \, (\varepsilon^2 (1 - \eta) + \eta)^2
}\quad  & \text{if} \quad \eta < \frac{\varepsilon^2}{1-\varepsilon^2} \\[6pt]
1+N & \text{if} \quad \eta = \frac{\varepsilon^2}{1-\varepsilon^2}
\end{cases}
\quad \textrm{with} \quad N=N^\text{Sq}+N^\text{Coh}.
    \label{ratiocomplex}
\end{align}
\end{widetext}
The maximum quantum advantage, $R$, is negatively impacted by increased $\eta$ and $\varepsilon$. Crucially, even small variations in $\varepsilon$ cause significant drops in R when operating in low-noise conditions, see Fig. [\ref{fig4}]. For completeness in Appendix~\ref{sec:appG} we show the QFI of the quantum strategy.

\section{Conclusions}\label{sec:conclusions}
We have shown that the rotational Doppler effect naturally arises as a consequence of the non-stationary boundary conditions imposed by a rotating surface. We have started by formally deriving the frequency shift from the Helmholtz equation with the boundary conditions, filling a gap in the literature. Making use of this result, we performed the quantization on two stationary Cauchy surfaces and derived the Bogoliubov transformation produced by the rotating surface. Employing this framework, we investigated the estimation of rotation frequency using the continuous-variable formalism. In particular, we proposed experimentally feasible states based on at least two Laguerre–Gaussian modes, one with single-mode squeezing and the other with displacement. We have demonstrated that employing homodyne detection on one of the scattered modes, this protocol achieves Heisenberg scaling in the noiseless regime. In a noisy scenario, the Heisenberg scaling is lost for bright beams, keeping a constant quantum advantage. To illustrate the generality of the approach, we studied two limiting cases of rotating surfaces, representing extreme situations: A metasurface that induces a definite change in the orbital angular momentum, and a surface with small random defects whose reflection produces an infinite set of different Doppler-shifted modes. Importantly, we emphasize that the proposed probe states and measurement are feasible with current technology, and our proposal can therefore be implemented with existing experimental capabilities. These results pave the way for advances in gyroscope technologies and, more broadly, in quantum-enhanced sensing.

\begin{acknowledgments}

We acknowledge fruitful discussions with Maximilian Reichert, Quimey Pears Stefano, and Miriam Lazo Moreno.

We acknowledge financial support from OpenSuperQ+100 (Grant No. 101113946) of the EU Flagship on Quantum Technologies,
from Project Grant No. PID2024-156808NB-I00 
and Spanish Ramón y Cajal Grant No. RYC-2020-030503-I funded by MICIU/AEI/10.13039/501100011033 and by “ERDF A way of making Europe” and “ERDF Invest in your Future”,
from the Spanish Ministry for Digital Transformation and of Civil Service of the Spanish Government through the QUANTUM ENIA project call-Quantum Spain, and by the EU through the Recovery, Transformation and Resilience Plan--NextGenerationEU within the framework of the Digital Spain 2026 Agenda, 
and from Basque Government through Grant No. IT1470-22, the IKUR Strategy under the collaboration agreement between Ikerbasque Foundation and BCAM on behalf of the Department of Education of the Basque Government,
and the Elkartek project KUBIBIT - kuantikaren berrikuntzarako
ibilbide teknologikoak (ELKARTEK25/79).  We acknowledge QTEP-CSIC via PTI-01 Quantum Technologies, and Project PID2022-143268NB-I00 funded by the Spanish Ministry of Science and Innovation (MCIN).

\end{acknowledgments}
\bibliography{apssamp}
\newpage 
 \onecolumngrid
\appendix
\newpage
\begin{section}{The paraxial approximation and Laguerre-Gaussian beams}\label{sec:app}
Laguerre-Gaussian beams are solutions of the vector Helmholtz equation $\left(\nabla^2+\omega^2/c^2\right)\vec{U}=0$ in the paraxial approximation. As we will study propagation in the absence of charges, the choice of  either the Couloumb or the transverse or the Weyl gauges leads to the electric potential being zero, and the vector Helmholtz equation for a monochromatic beam can equally well be studied for the electric field or the potential vector, since $\vec{E}=-i\omega\vec{A}$. At any rate, consider solutions $\vec{U}$ of the vector Helmholtz equation with an axis $z$ and an associated wavenumber $k$,
\begin{align}
    \vec{U}(\vec{r},t)= \vec{\epsilon}\:u(\vec{r})e^{-i(k z-\omega t)} ,
\end{align}
where $\vec{\epsilon}$ is a constant homogeneous unit vector field orthogonal to the $z$ axis. The paraxial approximation consists on discarding $\partial_z^2u$ in the vector Helmholtz equation, now scalar, leading to the paraxial equation
\begin{align}
    i\partial_z u = \frac{1}{2k}\nabla^2_\perp u+\frac{1}{2k}\left(\frac{\omega^2}{c^2}-k^2\right)\,.
\end{align}
This equation is formally equivalent to the time dependent Schr\"odinger equation, where now $z$ acquires the character of time, for the Hamiltonian
\begin{align}
    H_{\text{paraxial}}= -\frac{1}{2k}\vec{p}^2_\perp +C(k)\,,
\end{align}
describing a (negative mass) free particle in two dimensions, shifted by a constant. The general (formal) idea of spatial paraxial modes is to select a basis of functions of $x$ and $y$, $\left\{u^0_\alpha(x,y)\right\}_{\alpha\in I}$, with $I$ the corresponding index set,  and evolve the elements of that basis with $H_{\text{paraxial}}$ in $z$, such that 
\begin{align}
    u_\alpha(\vec{r})= e^{-i z H_{\text{paraxial}}}u^0_\alpha(x,y)\,.
\end{align}
Assuming the basis to be an orthonormal one for $L^2(\mathbb{R}^2)$, this guarantees orthonormality for fixed $z$, as $H_{\text{paraxial}}$ is essentially self-adjoint and the transformation from $u^0_\alpha$ to $u_\alpha(\bullet,z)$ is unitary. As presented here, \emph{any} basis for the transversal direction is amenable to this treatment. However, in order to comply with the approximation and to describe beam optics only those bases in which the dispersion around the axis is well controlled will be useful. In particular, different eigenbases of 2D harmonic oscillator Hamiltonians, with minimum on the axis, will be most relevant. These give rise to the \emph{Gaussian} beam modes, as the eigenbases will be  a Gaussian function multiplied by polynomial functions. This form, with dilations, is conserved for all $z$ because a harmonic oscillator Hamiltonian transforms under the action of the unitary $\exp(-i z H_{\text{paraxial}})$  into another positive definite quadratic Hamiltonian. In most textbooks the initial Hamiltonian is the isotropic 2D harmonic oscillator
\begin{align}
    H=\frac{1}{2}\vec{p}^2_\perp+ \frac{2r^2}{w_0^4}\,,
\end{align}
with $w_0$ being the waist radius, located at $z=0$. 

There are many instances in the literature noticing that under free particle evolution the eigenstates of a harmonic oscillator acquire both a time dependent frequency and a time rescaling. For one, see \cite{Takagi1990}. The connection to Gaussian beams has also been noted in the past, see for example \cite{Guerrero_2011}.
Here we want to stress two aspects that are not as present in previous literature. First, and almost trivially, that in principle the central wavenumber need not be that of the monochromatic beam. Secondly, that this perspective will enable us to assess the validity of the paraxial approximation.

For the sake of illustration, consider first the normalized ground state, 
\begin{align}\label{eq:u0ILE}
    u_0(x,y,0)=\frac{1}{w_0}\sqrt{\frac{2}{\pi}}\,e^{-r^2/w_0^2}\,.
\end{align}
This can be evolved using $H_{\text{paraxial}}$, yielding
\begin{align}
    u_0(x,y,z)= \frac{1}{w(z)}\sqrt{\frac{2}{\pi}}\,e^{-r^2/w^2(z)}e^{iz r^2/z_Rw^2(z)} e^{i\arctan(z/z_R)} e^{-i C(k) z}\,,
\end{align}
where $z_R=k w_0^2/2$ is the Rayleigh range determined by the waist radius $w_0$ and the chosen wavenumber $k$, and $w(z)=w_0\sqrt{1+(z/z_R)^2}$. Importantly, in order to compare with the standard expressions in the literature, there is a phase due to $C(k)$, that measures the difference between $\omega/c$ and the chosen axial wavenumber.
We shall consider the validity of the paraxial approximation in general and in this particular example for illustration. Namely, we have to compare the $\partial_z^2u$ term with $2k H_{\text{paraxial}}u$, and the approximation will be valid if, in some sense, it is small. For solutions of the paraxial equation, we have $i\partial_zu= H_{\text{paraxial}}u$, whence it follows that $\partial_z^2u= -H_{\text{paraxial}}^2u$. Thus, we have to estimate the paraxiality ratio 
\begin{align}
    \varepsilon_P=\frac{\left|\langle u|H_{\text{paraxial}}^2|u\rangle \right|}{2k\left|\langle u|H_{\text{paraxial}}|u\rangle\right|}\,,
\end{align}
and the approximation is valid if $\varepsilon_P\ll1$.
This is $z$-independent, and on eigenvectors of $H$ at $z=0$ with eigenenergy $E$ we can estimate it by computing
\begin{align}
    \left\langle u(0)\mid H_{\text{paraxial}}^n\mid u(0)\right\rangle = \left\langle u(0)\mid \left(C(k)-\frac{E}{k}+\frac{2r^2}{k w_0^4}\right)^n\mid u(0)\right\rangle \,.
\end{align}
For the initial function \eqref{eq:u0ILE} we have $E=2/w_0^2$, and
\begin{align}
    \varepsilon_P = \left|\frac{1}{k^2w_0^2}-\frac{C(k)^2}{2k(C(k)-1/kw_0^2)}\right|\,.
\end{align}
If $k=\omega/c$ the paraxial condition is therefore $kw_0\gg1$. A sufficient condition for paraxiality is
\begin{align}\label{eq:sufficient}
    \left|\left(\frac{\omega}{ck}\right)^2-1\right|\ll\frac{1}{k^2w_0^2}\ll 1\,.
\end{align}
We are interested in providing a sufficient condition for the validity of the paraxial approximation under rotation. As shown in the text, we consider modes with $k=(\omega+l\Omega)/c$ for a given $\omega$.  The description above qualitatively applies, and the paraxial approximation holds, if
\begin{align}\label{eq:shiftconstraint}
    \frac{4l\Omega}{\omega}\ll \frac{1}{k z_R}\ll 1\,.
\end{align}
In fact  $\langle r^2\rangle$ and $\langle r^4\rangle$ would have to be computed for other Laguerre-Gaussian modes, which for $k=\omega/c$  and dominant plane wave being $\exp(ikz)$ are usually presented as 
\begin{align}
    u_{l,p}(r,\phi,z)=C_{lp}^{LP}\frac{1}{w(z)}\left(\frac{r\sqrt{2}}{w(z)}\right)^{|l|}\exp\left( -\frac{r^2}{w^2(z)}\right)L_p^{|l|}\left(\frac{2r^2}{w^2(z)}\right)\exp(-ik\frac{r^2}{2R(z)})\exp(-il\phi)\exp(ig(z))
    \label{laguerre}
\end{align}
where $C_ {lp}^{LP}$ is a normalization constant, $L_p^l$ are the generalized Laguerre polynomials, $g(z)$ is the Gouy phase, $R(z)$ is the radius of curvature, $w(z)$ is the radius at which the intensity falls $1/e^2$ in the $z$ direction. The central $z$ wavenumber, $k$, is implicit in this notation, as is frequent in the literature. For definiteness, we collect here all the definitions,
\begin{align*}
    C_ {lp}^{LP}&=\sqrt{\frac{2p!}{\pi(p+|l|)!}}\\
    R(z)&=z\left(1+\left(\frac{z_R}{z}\right)^2\right)\\
    z_R&=\frac{n k w_0^2}{2}=\frac{\pi w_0^2 n}{\lambda}\\
    w(z)&=w_0\sqrt{1+\left(\frac{z}{z_R}\right)^2}\\
    g(z)&=\arctan\left(\frac{z_R}{z}\right)(N+1)
\end{align*}
where $n$ is the refractive index, $w_0$ is the waist radius and $z_R$  the Rayleigh range as above, and $N=|l|+2p$.  The qualitative conclusion of Eq.~\eqref{eq:shiftconstraint} holds.
Finally, we note that the LG basis presented in Eq.~\eqref{laguerre} depends on the waist radius $w_0$, this dependence will not play any role in our work so we will not make it explicit.

\subsubsection*{Polarization of the reflected beam}\label{app:polareq}
In the previous appendix we have examined the conditions of validity of the paraxial approximation. Here, in the context of that approximation, we want to analyze the polarization of a scattered wave when the incident one has homogeneous and constant polarization, for perfect conductor boundary conditions. The conclusion will be that we will be able to express reflective scattering, under some conditions for the scattering surface, in terms of a scalar wave.

Therefore we set out from the metallic boundary condition for incident electric field of constant and homogeneous polarization, 
\begin{align*}
    \vec{E}^\text{in}(\vec{r},t)= \bm{\epsilon}\, E^\text{in}(\vec{r},t)\,.
\end{align*}
To avoid confusion, let us denote here with $\bm{\epsilon}_2$ a polarization vector orthonormal to $\bm{\epsilon}$. Generically, the scattered electric field will present a superposition of those polarization states, in the form 
\begin{align*}
    \vec{E}^\text{out}(\vec{r},t) = \bm{\epsilon}\, E^\text{out}(\vec{r},t)+ \bm{\epsilon}_2\, E^\text{out}_2(\vec{r},t)\,.
\end{align*}
We want to study the boundary condition on a perfectly conducting surface determined  by $ f(\vec{r}_\text{S})=0$ or equivalently $z=h(x,y)$. Denote its normal vector field as  $\vec{n}(\vec{r}_\text{S})$, or more compactly $\vec{n}$. Thus, the perpendicular and parallel components of the polarization vectors $\bm{\epsilon}$ and $\bm{\epsilon}_2$ are at each point of the surface
\begin{align}
    \bm{\epsilon}_\perp(\vec{r}_\text{S}) &= \left(\bm{\epsilon}\cdot\vec{n}\right)\,\vec{n}\,,\qquad \bm{\epsilon}_\parallel(\vec{r}_\text{S}) = \bm{\epsilon}-\bm{\epsilon}_\perp(\vec{r}_\text{S})\,,\\ 
    \bm{\epsilon}_{2\perp}(\vec{r}_\text{S}) &= \left(\bm{\epsilon}_2\cdot\vec{n}\right)\,\vec{n}\,,\qquad \bm{\epsilon}_{2\parallel}(\vec{r}_\text{S}) = \bm{\epsilon}_2-\bm{\epsilon}_{2\perp}(\vec{r}_\text{S})\,.
\end{align}
The metallic boundary condition becomes
\begin{align}
    \bm{\epsilon}_\parallel\,\left(E^\text{in}+E^\text{out}\right) + \bm{\epsilon}_{2\parallel}\,E^\text{out}_2=0\,.
\end{align}
at each point of the surface. This vector equation has two components, that can be chosen as that in the direction $\bm{\epsilon}^\parallel$ and the one in the orthogonal direction $\vec{n}\times\bm{\epsilon}$. In other words,
\begin{align}
    \left[1-\left(\bm{\epsilon}\cdot\vec{n}\right)^2\right]\left(E^\text{in}+ E^\text{out}\right)-\left(\bm{\epsilon}\cdot\vec{n}\right)\left(\bm{\epsilon}_2\cdot\vec{n}\right)\,E^\text{out}_2&=0\,,\\
    \vec{n}\cdot\left(\bm{\epsilon}\times\bm{\epsilon}_2\right)\,E^\text{out}_2 &=0\,.
\end{align}
We are considering surfaces of limited roughness, and quasinormal incidence. It follows that, generically, $\vec{n}\cdot\left(\bm{\epsilon}\times\bm{\epsilon}_2\right)\neq0$ and $1-\left(\bm{\epsilon}\cdot\vec{n}\right)^2\neq0$. It follows that the boundary condition, a vector equation, decomposes generically into 
\begin{align}
    E^\text{in}(\vec{r}_\text{S})+E^\text{out}(\vec{r}_\text{S})&=0\,,\\
    E^\text{out}_2(\vec{r}_\text{S})&=0\,.
\end{align}
For the vector Helmholtz equation $(\nabla^2+\omega^2/c^2)\vec{A}=0$, in the $\phi=0$ gauge and a monochromatic wave, this entails $E_2^\text{out}=0$ elsewhere, not just on the surface, and we are limited to one scalar equation.

\section{Temporal modes and mode mapping}\label{app:temporal} 
Every field in the paraxial approximation can be decomposed, as mentioned above, as a sum of Laguerre-Gaussian modes, for which a definite orbital angular momentum $l$ is apparent \cite{yao2011orbital}. Yet, monochromatic beams are an unrealistic description of experimental procedures.
Instead, finite time-width beams are more feasible in the lab. Here we will use temporal Hermite-Gaussian (HG) packets, namely
\begin{align}
     \Phi_n(t; \tau_0, \omega_0, \theta_0, \sigma)
\equiv \frac{ \sqrt{\sigma} \sqrt{ \frac{2^{-n}}{n!} } }{ \sqrt[4]{2\pi} }
H_n\left( \frac{ \sigma(t-\tau_0) }{ \sqrt{2} } \right)
e^{ -\frac{1}{4} \sigma^2 (t - \tau_0)^2 }
e^{ -i (\omega_0 t + \theta_0) }
\end{align}
This set of functions, parameterised with the central frequency and time $\omega_0$ and $\tau_0$, the initial phase $\theta_0$ and, crucially, the frequency spread $\sigma$, and indexed by $n\in\mathbb{N}$, is orthonormal with respect to integration in time. They have to be put together with spatial modes in order to have a useful description of paraxial time packets, with a comoving waist through $g(t-z/c) u(\mathbf{r})\exp(-i k z)$ \cite{wunscheQuantizationGaussHermite2004}. Thus
the basis employed through this article is
\begin{align}
    \psi_{n,l,p}(t,r,\phi,z;\omega_0)= \Phi_n(t - z/c; \tau_0, \omega_0, \theta_0, \sigma)\cdot u_{l,p}(z,r,\phi)\,. \label{eq:B2}
\end{align}
Note that in the lhs of Eq.~\eqref{eq:B2} we have omitted some parameters. The dependence with all the parameters will be explicit only when it is relevant for the work.
Please observe also that we have omitted the explicit mention of the central longitudinal wavenumber $k$ for the spatial component.  For the analysis of the validity of the approximation we can take it to coincide with $\omega_0/c$. Then we have an additional condition, now involving the spread in frequency space of the (Fourier transformed) temporal function, 
\begin{align}\label{eq:sigmacondition}
    \sigma\ll\frac{c}{k w_0^2}\,.
\end{align}
Under this condition and the previously established ones for the spatial modes, 
the envelope of the wave pulse varies slowly in time and in the transverse coordinates. I.e, the rate of change in time, multiplied by the central period, and the rate of change in transverse coordinates, multiplied by the central wavelength, are both small. Were the central longitudinal wavenumber $k$ different from $\omega_0/c$, the analysis above, with minor modifications, shows that the paraxial conditions will hold for the Hermite-Gauss temporal/Laguerre-Gauss spatial modes under conditions \eqref{eq:sufficient} and \eqref{eq:sigmacondition}.

Following the analysis above, we now consider single-mode incidence on the rough surface, now Hermite-Gaussian in time and Laguerre-Gaussian in transversal coordinates. The reanalysis of  Eq.\eqref{eq:Ein} and Eq.\eqref{eq:Eout} produces 
\begin{align}
   &\tilde{E}^{ in}_{n,l,p}= \psi_{n,l,p}(t, r,\phi,z;\omega^\text{in})+c.c.\\&\tilde{E}_{n,l,p}^\text{out}= \sum_{l^\prime,p^\prime} c_{l^\prime,p^\prime;l , p }\:\psi_{n,l^\prime,p^\prime}(t, r,\phi,z;\omega^\text{in}-\Delta l\Omega)+c.c
     \label{eq:A3}
\end{align}
where  $\omega^\text{in}$ denotes the central frequency  of the basis used to describe the input beam. Notice that the expression for the scattered field involves a set of outgoing central temporal frequencies, $\omega^\text{in}-\Omega \Delta l$. In other words, it is not presented as a sum over one single orthonormal basis.  To further our goals, we have to rewrite the scattered field in such a way, as an expansion in one basis. One possible way to achieve this goal is to choose a reference outgoing frequency  $\omega^{\text{out}}$, and expand the temporal functions in the corresponding temporal Hermite-Gauss modes, 
using  properties of the Hermite polynomials \cite[7.388,~Eqs.~(6)~and~(7)]{gradshteyn2014table} (or alternatively \cite[\href{https://dlmf.nist.gov/18.17.E25}{(18.17.25)}]{NIST:DLMF} and \cite[\href{https://dlmf.nist.gov/18.17.E26}{(18.17.26)}]{NIST:DLMF}), leading to
\begin{align}
    \Phi_n(t; \tau_0, \omega^\text{in}-\Delta l \Omega, \theta_0^\text{in}, \sigma)&=\sum_m K_{m;n}(\beta_{l,l^\prime}) \Phi_m(t; \tau_0, \omega^{\text{out}}, \theta_0^\text{out}, \sigma),\label{eq:B4}\\
K_{m;n}(\beta_{l,l^\prime})&=\frac{ (-i)^{n-m}  }{ 2 }   \sqrt{\frac{m!}{n!}} (2\beta_{l,l^\prime})^{n-m} e^{-\frac{\beta_{l,l^\prime}^2}{2}} L_n^{n-m}(\alpha^2),\quad \beta_{l,l^\prime}:=\frac{\omega^\text{in}-\Delta l\Omega-\omega^{\text{out}}}{\sigma}\,.\label{eq:B5}
\end{align}
where $ \theta_0^{\text{out}}=(\omega^\text{in}-\omega^\text{out})\tau_0+\theta^\text{in}$ and $ n-m\geq0$. (More details of this calculation can be found in Appendix \ref{app:Hermiteoverlap}.)
We now rewrite Eq. \eqref{eq:A3} as
\begin{align}
\tilde{E}_{n,l,p}^\text{out}= \sum_{l^\prime, p^\prime, n^\prime} c_{l^\prime,p^\prime;l, p } K_{n^\prime;n}(\beta_{l^\prime,l }) \psi_{n^\prime,l^\prime,p^\prime}(t,r,\phi,z;\omega)+c.c..
      \label{A6}
\end{align}
Note that now the dependence with $\Omega$ appears both  in the basis, through $\omega$, and in the coefficients $K_{n^\prime;n}(\alpha_{l^\prime,l })$.\\

In our protocol we assume that there is a good prior value $\Omega_0$ for the rotational frequency $\Omega$ we want to determine, and we make use of that fact to further develop the expansions. Our analysis is complicated by the fact that the $\Omega$ dependence is present both in the coefficients and in the basis itself.

By using another interesting property of  Hermite-Gaussian temporal modes, namely
\begin{align}
    \partial_\omega \Phi_n(t; \tau_0, \omega, \theta_0, \sigma)=& -\,il \frac{\sqrt{2}}{\sigma}
\biggl(
  \sqrt{\frac{n}{2}}\,\Phi_{n-1}(t; \tau_0, \omega, \theta_0, \sigma)
  \;+\;
  \sqrt{\frac{n+1}{2}}\,\Phi_{n+1}(t; \tau_0, \omega, \theta_0, \sigma)
\biggr)\nonumber\\
&-
i\,\tau l\,\Phi_n(t; \tau_0, \omega, \theta_0, \sigma),
\label{modesd}
\end{align}
we can approximate expression \eqref{A6} for $\Omega$ close to a reference value $\Omega_0$. If we make a redefinition $ \omega^\text{out}\rightarrow \omega^\text{out}+\Delta l (\Omega-\Omega_0)\equiv \omega^\text{out}+\Delta l \:\delta\Omega$ we get 
\begin{align}
    &\Phi_n(t; \tau_0, \omega^\text{in}-\Delta l \Omega, \theta_0, \sigma)=\sum_m K_{m;n}(\beta_{l,l^\prime}) \Phi_m(t; \tau_0,\omega^\text{out}+ \Delta l\:\delta \Omega, \theta_0, \sigma),\\&\approx\sum_m K_{m;n}(\beta_{l,l^\prime})[\nu_{l^\prime} \Phi_{m-1}(t; \tau_0, \omega^\text{out}, \theta_0, \sigma)+\mu_{l^\prime}\Phi_{m}(t; \tau_0, \omega^\text{out}, \theta_0, \sigma)+\gamma_{l^\prime}\Phi_{m+1}(t; \tau_0, \omega^\text{out}, \theta_0, \sigma)] ,\label{A10}
\end{align}
where we have defined 
\begin{align*}
 \nu_{l^\prime}&= -i \frac{\sqrt{2}}{\sigma}\sqrt{\frac{n}{2}}\Delta l \delta\Omega, \\
 \mu_{l^\prime} &=-i\tau\Delta l \delta\Omega, \\
  \gamma_{l^\prime} &= -i \frac{\sqrt{2}}{\sigma}\sqrt{\frac{n+1}{2}}\Delta l \delta\Omega.
\end{align*}
We conclude that under the assumption $ \delta \Omega\ll1$ we can rewrite Eq. \eqref{A6} as: 
\begin{align}
\tilde{E}_{n,l,p}^\text{out}=\sum_{l^\prime, p^\prime, n^\prime} c_{l^\prime,p^\prime;l , p } [\nu_{l^\prime}K_{n^\prime+1;n}&(\beta_{l^\prime,l })+ (1+\mu_{l^\prime})K_{n^\prime;n}(\beta_{l^\prime,l })+\gamma_{l^\prime}K_{n^\prime-1;n}(\beta_{l^\prime,l })] \psi_{n^\prime,l^\prime,p^\prime}(t,r,\phi,z;\omega^\text{out})+c.c.\nonumber
      \label{A11}
\end{align}
We note that this transformation is not unitary in general, due to the first order approximation made in \eqref{A10}. Nevertheless, since the Fisher information only depends on first derivatives, the first order approximation is enough for our purposes \cite{sorelli2024gaussian}.
Using the notation introduced in the main text the mode transformation can be written in matrix form as follows 
\begin{align}
\psi_{\vec{m}}^\text{out}=\sum_{\vec{j},\vec{i}} \tilde U_{\vec{m}\vec{j}}\tilde G_{\vec{j}\vec{i}} \psi_\vec{i}^\text{in} ,\quad \forall \:\vec{m}
\end{align}
where given that $\vec{i}=\lbrace n^\prime,l^\prime,p^\prime \rbrace $ and $\vec{j}=\lbrace n,l , p \rbrace $  then,
\begin{align*}
\tilde G_{\vec{i}\vec{j}}&=\delta_{l^\prime;l } \delta_{p^\prime;p }\left[\nu_{l^\prime}\delta_{n^\prime-1;n}+(1+\mu_{l^\prime})\delta_{n^\prime;n}+\gamma_{l^\prime} \delta_{n^\prime+1;n}\right],\\
\tilde U_{\vec{i}\vec{j}}&=K_{n^\prime;n}\:c_{l^\prime,p^\prime;l,  p }\:.
\end{align*}

\subsection{Computing Eqs.~\eqref{eq:B4} and ~\eqref{eq:B5} }\label{app:Hermiteoverlap}
We compute
\begin{align*}
    K_{m;n}(\omega^\text{in},\theta_0^\text{in},\omega^\text{out},\theta_0^\text{out})&=\int \Phi_n(t; \tau_0, \omega^\text{in}, \theta_0^\text{in}, \sigma)\Phi_m^*(t; \tau_0, \omega^\text{out}, \theta_0^\text{out}, \sigma)   \hspace{1cm} m\leq n,  
\end{align*}
where
\begin{align*}
     \Phi_n(t; \tau_0, \omega_0, \theta_0, \sigma)
\equiv \frac{ \sqrt{\sigma} \sqrt{ \frac{2^{-n}}{n!} } }{ \sqrt[4]{2\pi} }
H_n\left( \frac{ \sigma(t-\tau_0) }{ \sqrt{2} } \right)
e^{ -\frac{1}{4} \sigma^2 (t - \tau_0)^2 }
e^{ -i (\omega_0 t + \theta_0) }.
\end{align*}
Substituting,
\begin{align*}
    K_{m;n}&= \frac{ \sqrt{\sigma} \sqrt{ \frac{2^{-n}}{n!} } }{ \sqrt[4]{2\pi} }  \frac{ \sqrt{\sigma} \sqrt{ \frac{2^{-m}}{m!} } }{ \sqrt[4]{2\pi} }\int H_n\left( \frac{ \sigma(t-\tau_0) }{ \sqrt{2} } \right)H_m\left( \frac{ \sigma(t-\tau_0) }{ \sqrt{2} } \right)
e^{ -\frac{1}{2} \sigma^2 (t - \tau_0)^2 }
e^{ -i ((\omega^\text{in}-\omega^\text{out})t + \theta_0^\text{in}-\theta_0^\text{out}) }.
\end{align*}
Defining $x=\frac{ \sigma(t-\tau_0) }{ \sqrt{2} }  $ , $\theta_0^{\prime\text{in}}= \omega^\text{in}\tau_0+\theta_0^\text{in}$, $\theta_0^{\prime\text{out}}= \omega^\text{out}\tau_0+\theta_0^\text{out}$  and using that $e^{-i\omega t}= \cos(\omega t)-i\sin(\omega t)$ we get,
\begin{align*}
    K_{m;n}&= \frac{ \sigma e^{ -i(\theta_0^{\prime\text{in}}-\theta_0^{\prime\text{out}})}  }{ \sqrt{2\pi} }  \sqrt{ \frac{2^{-n}}{n!} } \sqrt{ \frac{2^{-m}}{m!} } \frac{\sqrt{2}}{\sigma}\int H_n\left( x \right)H_m\left( x \right)
e^{ -x^2 }
e^{ -i (\omega^\text{in}-\omega^\text{out})\frac{\sqrt{2}x}{\sigma} }\\
 K_{m;n}&= \frac{  e^{ -i(\theta_0^{\prime\text{in}}-\theta_0^{\prime\text{out}})}  }{ \sqrt{\pi} }  \sqrt{ \frac{2^{-n}}{n!} } \sqrt{ \frac{2^{-m}}{m!} } \int H_n\left( x \right)H_m\left( x \right)
e^{ -x^2 }
\left(\cos((\omega^\text{in}-\omega^\text{out})\frac{\sqrt{2}x}{\sigma}) -i \sin((\omega^\text{in}-\omega^\text{out})\frac{\sqrt{2}x}{\sigma})\right). \\
\end{align*}
In the following we will study separately the cases $n-m$ being even or odd.
\subsubsection{$n-m$ is even}
If $n-m$ is even the product of the two Hermite polynomials is symmetric, whence  the sine contribution to the integral is zero and the integral reduces to
\begin{align*}
 K_{m;n}&= \frac{  e^{ -i(\theta_0^{\prime\text{in}}-\theta_0^{\prime\text{out}})}  }{ \sqrt{\pi} }  \sqrt{ \frac{2^{-n}}{n!} } \sqrt{ \frac{2^{-m}}{m!} } \int H_n\left( x \right)H_m\left( x \right)
e^{ -x^2 }
\cos((\omega^\text{in}-\omega^\text{out})\frac{\sqrt{2}x}{\sigma}) .
\end{align*}
Using \cite[7.388,~Eqs.~(6)~and~(7)]{gradshteyn2014table},
\begin{align*}
& K_{m;n}=\\ &\frac{  e^{ -i(\theta_0^{\prime\text{in}}-\theta_0^{\prime\text{out}})}  }{ \sqrt{\pi} }  \sqrt{ \frac{2^{-n}}{n!} } \sqrt{ \frac{2^{-m}}{m!} } 2^{n-1/2}\sqrt{\frac{\pi}{2}}m!(-1)^{(n-m)/2} \left(\frac{2(\omega^\text{in}-\omega^\text{out})^2
 }{\sigma^2}\right)^{(n-m)/2}L_n^{n-m}(\frac{(\omega^\text{in}-\omega^\text{out})^2}{\sigma^2}) e^{-\frac{(\omega^\text{in}-\omega^\text{out})^2}{ 2\sigma^2}}.
\end{align*}
Defining $\beta=\frac{\omega^\text{in}-\omega^\text{out}}{ \sigma}$ and simplifying we get,
\begin{align*}
& K_{m;n}=\\ &\frac{ e^{ -i(\theta_0^{\prime\text{in}}-\theta_0^{\prime\text{out}})}  }{ 2 }   \sqrt{ 2^{-(m-n)} }  \sqrt{\frac{m!}{n!}}(-1)^{(n-m)/2} \left(\sqrt{2}\beta\right)^{n-m}L_n^{n-m}(\beta^2) e^{-\frac{\beta^2}{ 2}}\\
&\frac{ e^{ -i(\theta_0^{\prime\text{in}}-\theta_0^{\prime\text{out}})}  }{ 2 }     \sqrt{\frac{m!}{n!}}(-1)^{(n-m)/2} \left(2\beta\right)^{n-m}L_n^{n-m}(\beta^2) e^{-\frac{\beta^2}{ 2}}
\end{align*}
\subsubsection{$n-m$ is odd}
If $n-m$ is odd the product of the two Hermite polynomials is antisymmetric meaning, so the cosine contribution is zero and the integral reduces to
\begin{align*}
 K_{m;n}&= -i\frac{  e^{ -i(\theta_0^{\prime\text{in}}-\theta_0^{\prime\text{out}})}  }{ \sqrt{\pi} }  \sqrt{ \frac{2^{-n}}{n!} } \sqrt{ \frac{2^{-m}}{m!} } \int H_n\left( x \right)H_m\left( x \right)
e^{ -x^2 } \sin((\omega^\text{in}-\omega^\text{out})\frac{\sqrt{2}x}{\sigma}) .
\end{align*}
Using \cite[7.388,~Eqs.~(6)~and~(7)]{gradshteyn2014table},
\begin{align*}
 K_{m;n}&= -i\frac{ e^{ -i(\theta_0^{\prime\text{in}}-\theta_0^{\prime\text{out}})}  }{ \sqrt{\pi} }  \sqrt{ \frac{2^{-n}}{n!} } \sqrt{ \frac{2^{-m}}{m!} } 2^{n-1} (-1)^{(n-m-1)/2}\sqrt{\pi} m!(\sqrt{2}\beta)^{n-m} e^{-\frac{\beta^2}{2}} L_n^{n-m}(\beta^2)\\
  K_{m;n}&= -i\frac{  e^{ -i(\theta_0^{\prime\text{in}}-\theta_0^{\prime\text{out}})}  }{ 2 }   \sqrt{2^{-(m+n)}}2^{n} (-1)^{(n-m-1)/2}\sqrt{\frac{m!}{n!}} (\sqrt{2}\beta)^{n-m} e^{-\frac{\beta^2}{2}} L_n^{n-m}(\beta^2)\\
  K_{m;n}&= -i\frac{ e^{ -i(\theta_0^{\prime\text{in}}-\theta_0^{\prime\text{out}})}  }{ 2 }   (-1)^{(n-m-1)/2}\sqrt{\frac{m!}{n!}} (2\beta)^{n-m} e^{-\frac{\beta^2}{2}} L_n^{n-m}(\beta^2)
\end{align*}
The odd and even cases can be written in a single formula,
\begin{align*}
  K_{m;n}&= (-i)^{n-m}\frac{ e^{ -i(\theta_0^{\prime\text{in}}-\theta_0^{\prime\text{out}})}  }{ 2 }   \sqrt{\frac{m!}{n!}} (2\beta)^{n-m} e^{-\frac{\beta^2}{2}} L_n^{n-m}(\beta^2)\,.
\end{align*}
With a suitable choixe of $ \theta_0^{\text{out}}$, namely ($ \theta_0^{\text{out}}=(\omega^\text{in}-\omega^\text{out})\tau_0+\theta^\text{in}$), this becomes
\begin{align*}
  K_{m;n}&= \frac{ (-i)^{n-m}  }{ 2 }   \sqrt{\frac{m!}{n!}} (2\beta)^{n-m} e^{-\frac{\beta^2}{2}} L_n^{n-m}(\beta^2)\,.
\end{align*}

\end{section}

\section{Light interaction model for a rough surface}\label{Roughmodel}

The reflection of a wavefront from a rough mirror introduces a position-dependent phase shift. The mirror's surface can be described by a height function $h(r, \phi)$, where $r$ and $\phi$ are polar coordinates. This height variation is modeled as a zero-mean, stationary random process. A key parameter is its standard deviation, $\sigma_h$, the root-mean-square (RMS) roughness.

An incoming electric field, $E_{\text{in}}(r, \phi)$, is transformed upon reflection to:
\begin{align}
    E_{\text{out}}(r, \phi,z=0) =-E_{\text{in}}(r, \phi,z=0) \, e^{i 2 k h(r, \phi)},
\end{align}
where $k=2\pi/\lambda$ is the wavenumber. For a specific surface realization $h(r, \phi)$, an incoming Laguerre-Gaussian (LG) mode $u_{l^\prime , p^\prime }(r, \phi)$ scatters into a superposition of outgoing modes. The coupling coefficients $c_{l,p;l^\prime , p^\prime }$ are given by the overlap integral:
\begin{align}
 c_{l,p;l^\prime , p^\prime } = -\iint u_{l,p}^*(r, \phi,0) \, e^{i 2 k h(r, \phi)} \, u_{l^\prime,  p^\prime }(r, \phi,0) \, r \, \diff r \, \diff \phi.
\end{align}
To simplify the notation in the following $u_{l^\prime,  p^\prime }(r, \phi,0)=u_{l^\prime,  p^\prime }(r, \phi)$.
We now consider the weak scattering approximation, which is valid when the RMS roughness is much smaller than the wavelength ($k \sigma_h \ll 1$). In this regime, the phase term can be expanded to first order, $\exp{2i k h(r,\phi)}\approx 1+ 2i k h(r,\phi)$, leading to 
\begin{align}
    c_{l,p;l^\prime , p^\prime } \approx- \delta_{l,l^\prime }\delta_{p, p^\prime } -2 i k \iint u_{l,p}^*(r, \phi) \, h(r, \phi) \, u_{l^\prime , p^\prime }(r, \phi) \, r \, \diff r \, \diff \phi. 
\end{align}
The first term represents the specular reflection, while the second integral, which we denote $ih_{lp;l^\prime  p^\prime }$, describes the scattering between different modes. Note that the  integral gives us the matrix elements in an orthonormal basis of an operator $\hat{h}$, that is hermitian, as it is diagonal in the position basis with eigenvalues $h(r,\phi)$. Thus
\begin{align*}
    \left(h_{lp;l^\prime  p^\prime }\right)^*&=\left(\iint u_{l,p}^*(r, \phi) \, h(r, \phi) \, u_{l^\prime , p^\prime }(r, \phi) \, r \, \diff r \, \diff \phi\right)^*=\iint u_{l,p}(r, \phi) \, h^*(r, \phi) \, u_{l^\prime , p^\prime }^*(r, \phi) \, r \, \diff r \, \diff \phi\\&=\iint u_{l,p}(r, \phi) \, h(r, \phi) \, u_{l^\prime , p^\prime }^*(r, \phi) \, r \, \diff r \, \diff \phi= h_{l^\prime p^\prime;l  p}.
\end{align*}
To derive a tractable functional form, we turn to the statistical properties of the surface. A standard model assumes the surface is a Gaussian random process with a Gaussian autocorrelation function \cite{goodman2015statistical}. Via the Wiener-Khinchin theorem, this implies that the Power Spectral Density (PSD) of the surface is also a Gaussian function of spatial frequency. The average power scattered between modes is directly related to this PSD. While the exact relationship is a complex integral, this physical link strongly motivates modeling the scattering amplitude with a Gaussian profile as a function of the change in mode indices.

Based on this, we propose the following scattering matrix:
\begin{align}
     c_{l,p;l^\prime , p^\prime } = (-\sqrt{1-\varepsilon^2}+i\frac{\varepsilon}{\mathcal{N}})\:\delta_{l,l^\prime }\delta_{p, p^\prime }-i\varepsilon \frac{ e^{-\frac{(l-l^\prime )^2}{4\sigma_l^2}} e^{-\frac{(p- p^\prime )^2}{4\sigma_p^2}}}{\mathcal{N}}e^{i\theta_{l, p;l^\prime ,p^\prime}}.
\end{align}
The parameter $\varepsilon^2 \ll 1$ represents the fraction of the amplitude of the incident mode scattered out, and is proportional to $k\sigma_h$. The terms $\sigma_l$ and $\sigma_p$ control the modal spread of the scattered light and are related to the surface correlation length and $\theta_{l,p;l^\prime ,p^\prime}$ is an antisymmetric phase $\theta_{l,p;l^\prime ,p^\prime} = -\theta_{l^\prime ,p^\prime;l,p}$ due to $\hat{h} $ being hermitian. The normalization constant $\mathcal{N}$ is chosen such that the transformation is normalized,
\begin{align*}
    \mathcal{N}^2=(\sqrt{2\pi}|\sigma_l|\theta_3(\pi,e^{-2\pi^2\sigma_l^2})-1)(\sqrt{2\pi}|\sigma_p|\theta_3(\pi,e^{-2\pi^2\sigma_p^2})-1),
\end{align*}
where $\theta_3(u,q)$ is the Jacobi theta function defined as $\theta_3(u,q)=1+2\sum_{n=1}^\infty q^{n^2}\cos(2\: n\: u)$.

\begin{section}{Continuous variables conventions}\label{CV}
The usual way of quantizing a non-relativistic dynamical, bosonic, degree of freedom starts from considering two self-adjoint operators $\hat{x}$ and $\hat{p}$, called `position' and `momentum', and making them satisfy the canonical commutation relation (ccr):
\begin{equation}\label{eq:ccr}
    [\hat{x}, \hat{p}] = i \hbar \hat{\Id}.
\end{equation}
The $\hat{x}$ and $\hat{p}$ operators are unbounded operators exhibiting a continuous spectrum, hence the name `quantum continuous variables'. This means that in general, even a single, non-relativistic bosonic degree of freedom requires a (countably) infinite-dimensional basis for its Hilbert space \footnote{Infinite-dimensional representations of the ccr algebra do exist but in order to guarantee that there are indeed irreducible representations, one should use the exponentiated version of the ccr, a particular case of the Baker-Campbell-Hausdorff formula known as Weyl relation: $\exp(i\lambda \hat{x}) \exp(i\mu \hat{p}) =  \exp(i\mu \hat{p}) \exp(i\lambda \hat{x}) \exp(i\lambda\mu\hbar \hat{\Id})$ with $\lambda, \mu \in \reals$.}. This translates to infinite-dimensional matrix representations of the density operator, which motivates the use of other representations via quasiprobability distributions, like the Wigner function \cite{serafini2023quantum}. 

For the `field operators' $\hat{a}^\dagger$ and $\hat{a}$ we follow the conventions of Ref.~\cite{pinel2013quantum}, and references therein: 
\begin{align}
    \hat{a} &:= \frac{1}{{2}}(\hat{x}+i\hat{p}),\\
    \hat{a}^\dagger &:= \frac{1}{{2}}(\hat{x}-i\hat{p}).
\end{align}
Their commutation relation is $[\hat{a}, \hat{a}^\dagger] = \frac{\hbar}{2}\hat{\Id} $. Hence, making $\hbar = 2$ is natural in order to keep $[\hat{a}, \hat{a}^\dagger] = \hat{\Id}$. Note that $\hat{x} = \hat{a}^{\dagger} + \hat{a}, \quad \hat{p} = i(\hat{a}^{\dagger} - \hat{a})$.

In the multimode situation, whether one uses position and momentum, or creation and annihilation operators, defines a choice of basis. In general we have
\begin{equation}\label{eq:modes-commutator}
    [\hat{a}_\vec{m}(\omega), \hat{a}_{\vec{m}^\prime}(\omega^\prime)^\dagger] = \hat{\Id} \delta_{\vec{m},\vec{m}^\prime} \delta(\omega-\omega^\prime).
\end{equation}
In our discussion, a `mode' is the combination of discrete indices in vector form $\vec{m} = (m_1, \ldots, m_k)^\trans$,  while frequency $\omega$ is considered a parameter. Note that in the main text we typically specify a mode with three indices $\vec{m} = \lbrace n,l,p \rbrace$, but here we keep the discussion more general. It is important to bear in mind that all $m_k$ belong to countable sets, such that vectors $\vec{m} = (m_1, \ldots, m_k)^\trans$ are also countable. This is indeed the case for our setup, since all $m_k \in \integers$. In the following we set the cardinality of the modal set to $N$. In Eq.~\eqref{eq:modes-commutator}, the vectorial Kronecker delta expands in components as one should expect: $\delta_{\vec{m},\vec{m}^\prime} \equiv \prod_{i=1}^k \delta_{m_i,{m_i}^\prime}$.

Throughout this paper we use the so-called `real basis', defined by: $\hat{\mathbf{r}}:= (\hat{x}_\vec{1},\hat{p}_\vec{1},\ldots,\hat{x}_\mathbf{d},\hat{p}_\mathbf{d})^\trans$. Note that we define $\hat{\vec{r}}_\vec{i} = (\hat{x}_\mathbf{i},\hat{p}_\mathbf{i})^\trans$. In this basis  the ccr are $\left[\hat{\mathbf{r}}, \hat{\mathbf{r}}^\intercal\right] = i \hat{\Omega}$, where $\hat{\Omega} = \bigoplus_{j=1}^N \hat{\Omega}_1$ is a real, symmetric, quadratic form \footnote{We write $\hat{\Omega}$  to distinguish the symplectic form from the angular velocity $\Omega$, a slight abuse of notation, since `hats' are reserved for operators and $\hat{\Omega}$ is a matrix.} (known as `real symplectic form'), and 
\begin{equation}
 \hat{\Omega}_1 =
    \begin{pmatrix}
       0 & 1\\
       -1 & 0
    \end{pmatrix}.
\end{equation} 

Within the bosonic CV quantum systems, quantum Gaussian states are defined as the ones arising from Hamiltonians that are at most quadratic in the field operators. This means that they can be fully described by their first and second moments, known as displacement vector and covariance matrix. The components associated to a given mode and pair of modes are, respectively:
\begin{align}
    {\mathbf{r}}_\vec{i} &= \langle \hat{\mathbf{r}}_\vec{i} \rangle \\ \left[\Sigma\right]_{\vec{i} \vec{j}} &= \frac{1}{2}\langle\hat{\mathbf{r}}_\vec{i} \hat{\mathbf{r}}^\trans_\vec{j} + \hat{\mathbf{r}}_\vec{j} \hat{\mathbf{r}}^\trans_\vec{i} \rangle - \langle \hat{\mathbf{r}}_\vec{i} \rangle \langle \hat{\mathbf{r}}_\vec{j}^\trans \rangle
\end{align}
   with $\vec{i}, \vec{j} \in[\mathbf{d}] \cong [N]  \equiv \lbrace 1, \ldots , N\rbrace $, and where expectation values are defined via $\langle \hat{A} \rangle = \Tr (\rho \hat{A})$, well-defined as long as the operator $\rho\hat{A}$ is trace-class. This is indeed the case for all of our examples. The Hilbert space associated to our quantum theory is obtained simply by combining: $$\hilbert = \otimes_{\vec{i}=\vec{1}}^\vec{d} \hilbert_{\vec{i}} \sim \otimes_{k=1}^N \hilbert_{m_k}.$$ We denote the vacuum of the total space as $\ket{0} \in \hilbert$.

    In the following we omit the modal labeling of the operators described, for simplicity. The displacement operator --whose name comes from a phase-space visualization of its action-- is defined as:
\begin{equation}
    \hat{D}(\alpha):= e^{\alpha \hat{a}^\dagger - \alpha^*\hat{a}}\equiv e^{-\abs{\alpha}^2/2}e^{\alpha \hat{a}^\dagger}e^{-\alpha^* \hat{a}},
\end{equation}
with $\alpha \in \complex$.
Its action on the vacuum produces a coherent state: $\hat{D}(\alpha)\ket{\alpha} = \ket{0}$. The set $\lbrace \ket{\alpha} \rbrace_{\alpha\in \complex}$ forms an overcomplete basis whose resolution of the identity is:
\begin{equation}
\hat{\Id} = \frac{1}{\pi}\int_\complex \diff^{2}\alpha |\alpha\rangle\langle\alpha |, 
\end{equation}
where $\diff^{2}\alpha \equiv \diff \left(\Re\alpha\right)\diff \left(\Im\alpha\right)$.
Coherent states can be also defined as the eigenstates of the annihilation operator: $\hat{a}\ket{\alpha} = \alpha\ket{\alpha} $. Implementation of the displacement operation on any state, pure or mixed, can be arbitrarily well approximated by mixing it with a highly excited coherent state at a beam splitter \cite{Paris1996}.
 
 The other relevant Gaussian operation in this paper  is squeezing. In its single-mode version, $\hat{S}(\xi)$ acts on the vacuum:
\begin{equation}
    \ket{\xi}\equiv \hat{S}(\xi)\ket{0} := \exp(-\xi\hat{a}^{\dagger\;2}+\xi^*{\hat{a}}^2)\ket{0},
\end{equation}
where $\xi = s e^{i\theta}$ is the squeezing parameter. For $\xi \in \reals $ the covariance matrix of a single-mode squeezed state is $\text{diag}(e^{-2s}, e^{2s})$.

\end{section}
\section{ Optimality of the classical strategy}\label{sec:appD}
In this Appendix we study when homodyne is optimal for classical strategies. 
For Gaussian states the quantum Fisher information can be fully expressed in terms of the mean displacement and covariance matrix. For single-mode Gaussian states with constant purity $P= \det(\Sigma_{\mathbf{m}})^{-1/2}=:|\Sigma_{\mathbf{m}}|^{-1/2}$ the QFI is reduced to \cite{pinel2013quantum,monras}:
\begin{align}
  J = \frac{1}{2} \frac{\text{Tr}[(\Sigma_{\mathbf{m}}^{-1}\Sigma_{\mathbf{m}}')^2]}{1+P} + (\mathbf{r}_{\mathbf{m}}') ^{\trans}\Sigma_{\mathbf{m}}^{-1}\mathbf{r}'_{\mathbf{m}} .
    \label{D1}
\end{align}
For the classical state
\begin{align}
|\Psi_\text{C}\rangle=\bigotimes_{\mathbf{d}\in I}\hat{D}_{\mathbf{d}}(\alpha_\mathbf{d})|0\rangle,
\end{align}
the mean vector and the covariance matrix are given by:
\begin{align*}
\mathbf{r}_{\mathbf{m}} 
=\sqrt{1-\eta}\sum_ \mathbf{i}\begin{pmatrix}
\Re[U_{\mathbf{m}\mathbf{i}} ]x_\mathbf{i}-\Im[U_{\mathbf{m}\mathbf{i}} ]p_\mathbf{i} \\
\Im[U_{\mathbf{m}\mathbf{i}} ]x_\mathbf{i}+\Re[U_{\mathbf{m}\mathbf{i}} ]p_\mathbf{i}
\end{pmatrix},\quad \Sigma_{\mathbf{m}}=
\begin{pmatrix}
 1&  0  \\
0 & 1
\end{pmatrix}.
\end{align*}
where the prefactor $ \sqrt{1-\eta}$ is due to noise, see [\ref{strategy}]. Since $\Sigma_{\mathbf{m}}'=0 $ only the second term in \eqref{D1} contributes. Computing the QFI we find:
\begin{align}
  J=(\mathbf{r}_{\mathbf{m}}') ^{\trans}\Sigma_{\mathbf{m}}^{-1}\mathbf{r}'_{\mathbf{m}}&=(1-\eta)\left(\sum_\mathbf{i}\Re[U'_{\mathbf{m}\mathbf{i}} ]x_\mathbf{i}-\Im[U'_{\mathbf{m}\mathbf{i}} ]p_\mathbf{i}\right)^2+\left(\sum_\mathbf{i}\Im[U'_{\mathbf{m}\mathbf{i}} ]x_\mathbf{i}+\Re[U'_{\mathbf{m}\mathbf{i}} ]p_\mathbf{i}\right)^2\nonumber\\&=4(1-\eta)(\sum_\mathbf{i}Re^2[U'_{\mathbf{m}\mathbf{i}}\alpha_\mathbf{i} ]+Im^2[U'_{\mathbf{m}\mathbf{i}}\alpha_\mathbf{i}])=4(1-\eta)\sum_\mathbf{i}|U'_{\mathbf{m}\mathbf{i}}|^2|\alpha_\mathbf{i} |^2
\end{align}
where we have used $ x_\mathbf{l}=2\Re[\alpha_l]$ and $ p_\mathbf{l}=2\Im[\alpha_l]$ in the last equality. Note that the quantum Fisher information is independent of the phase of the phases of $\alpha_i$.
In the following we show the Fisher information \eqref{CFisher} when the measurement considered is homodyne see [\ref{HOmodyne}]. 
\begin{equation}
    F_\text{C} = (1-\eta)(\sum_i\Re[U'_{\mathbf{m}\mathbf{i}} ]x_\mathbf{i}-\Im[U'_{\mathbf{m}\mathbf{i}} ]p_\mathbf{i})^2=4(1-\eta)(\sum_i\Re[U'_{\mathbf{m}\mathbf{i}}\alpha_i ])^2
\end{equation}
by defining $\alpha_ie^{i\phi_i}=|\alpha_i|$ choosing $\phi_i=-\operatorname{arg}\{U_{\mathbf{m}\mathbf{i}}' \}$ the classical Fisher information is,
\begin{equation}
    F_\text{C} = 4(1-\eta)\sum_i|U'_{\mathbf{m}\mathbf{i}}|^2|\alpha_i |^2
\end{equation}
Since this expression is the same as the quantum Fisher information we conclude that homodyne is an optimal measurement for the classical strategy.
Optimizing location of the energy among the different modes under the constriction $N=\sum_\mathbf{i}|\alpha_\mathbf{i}|^2$ we get
\begin{align}
  \max_{\{\alpha_\mathbf{i}\}}\{(\mathbf{r}_{\mathbf{m}}') ^{\trans}\Sigma_{\mathbf{m}}^{-1}\mathbf{r}'_{\mathbf{m}}\}=4(1-\eta)N |U_{\mathbf{m}\mathbf{l}}'|^2, \quad \text{ with } \quad |U_{\mathbf{m}\mathbf{l}}'|=\max_\mathbf{i}\{ U_{\mathbf{m}\mathbf{i}}'\}
\end{align}
and the optimal choice is given by $ \alpha_\mathbf{i}=0$, $\forall \mathbf{i}\neq \mathbf{l}$ and $|\alpha_\mathbf{l}|=\sqrt{N}$. So we get
\begin{align}
\operatorname{max}_{\phi}\{F\} =4(1-\eta) N|U_{\mathbf{m}\mathbf{l}}'|^2.
    \label{D5}
\end{align}
And consequently, the mean vector and covariance matrix of the optimal state are given by:
\begin{align*}
\mathbf{r'}_{\mathbf{m}} 
=\sqrt{1-\eta}\begin{pmatrix}
2\sqrt{N}|U'_{\mathbf{m}\mathbf{l}}|\\
0
\end{pmatrix},\quad \Sigma_{\mathbf{m}}=
\begin{pmatrix}
 1&  0  \\
0 & 1
\end{pmatrix}.
\end{align*}
As was stated in Section~[\ref{HOmodyne}] the position quadrature contains all the information of the parameter while the momentum operator does not contain any information, that is to say, its derivative with respect to the parameter is $0$.
As expected the Fisher information for the classical strategy follows the standard limit.\\
It is common in the literature to relate the Fisher information with some parameters of the beam. As is shown in \cite{reichert2024heisenberg} the Fisher information for coherent states can be written as,
\begin{align*}
    F=4\operatorname{Re}[\int \diff t \partial_\Omega s(t) (\partial_\Omega s(t))^* ]=4\operatorname{Re}[\int \diff t \partial_\Omega (\sum_{n^\prime, l^\prime, p^\prime} \alpha_{n^\prime, l^\prime, p^\prime}\:\psi_{n^\prime l^\prime p^\prime})\partial_\Omega (\sum_{n^\prime,  l^\prime,  p^\prime} \alpha_{n^\prime,  l^\prime, p^\prime}^*\:\psi_{n^\prime l^\prime p^\prime}^*) ]\\=4\operatorname{Re}[\int \diff t  (\sum_{n^\prime, l^\prime, p^\prime } \alpha_{n^\prime l^\prime p^\prime }\Delta l \:it \:\psi_{n^\prime l^\prime p^\prime}) (\sum_{n^\prime, l^\prime, p^\prime} \alpha_{n^\prime l^\prime p^\prime}^*\Delta l \:(-it)\:\psi_{n^\prime l^\prime p^\prime }^*) ]=4\operatorname{Re}[\int \diff t \: t^2 |F(t)|^2 ]
\end{align*}
where we have written the signal in the LG basis, $s(t)=\sum_{n^\prime l^\prime p^\prime }\alpha_{n^\prime l^\prime p^\prime}\:\psi_{n^\prime l^\prime p^\prime } $ and we have defined $F(t)=\sum_{n^\prime l^\prime p^\prime} \alpha_{n^\prime l^\prime p^\prime}\Delta l \: \:\psi_{n^\prime l^\prime p^\prime}$. It can be seen that the Fisher information is related with the time-width  of the function $F(t)$. Where $F(t)$ depends on the reflected signal $s(t)$. This ``pseudo-time-width'' plays the same role for the angular velocity estimation as the standard time-width for the frequency estimation (velocity estimation). For a fair comparison in the main text we consider classical and quantum strategies that employ the same modes so that they use the same $\Delta l$ so that the quantum strategy will never have a larger ``pseudo-time-width''.

\begin{section}{Fisher information of the surface models} \label{sec:appEx}
In this section we derive the expressions shown in the two examples studied. The two surfaces considered, --metasurface and complex surface--,
are not invariant under rotations, meaning that $ f(z,\phi,r)\neq f(z,\phi+\theta,r)$ for any $\theta$ so the reflected beam would experience an OAM shift and RDE could be used to estimate the rotation frequency.

\subsection{Metasurface}\label{app.Meta}
First we show that if the surface induces a definite change in the orbital angular momentum  $\Delta l^* $. Then, the general transformation 
\begin{align}
    \hat{a}^\text{out}_{n, l,  p }(\omega^\text{out})=&\sum_{l^\prime p^\prime n^\prime} c_{l^\prime,p^\prime;l ,p } [\nu_lK_{n^\prime+1;n}(\beta_{l^\prime,l })+ (1+\mu_l)K_{n^\prime;n}(\beta_{l^\prime,l })+\gamma_lK_{n^\prime-1;n}(\beta_{l^\prime,l })] \hat{a}^\text{in}_{n^\prime,l^\prime,p^\prime}\nonumber\\ &\quad\quad\quad\quad\quad\beta_{l,l^\prime}:=\frac{\omega^\text{in}-\Delta l\Omega-\omega^{\text{out}}}{\sigma}\,
\end{align}
can be rewritten as
\begin{align}
      \hat{a}^\text{out}_{n, l,  p}(\omega^\text{out})=\sum_{l^\prime p^\prime n^\prime} -\delta_{l^\prime-\Delta l^*,l }\delta_{p^\prime, p } [\nu_lK_{n^\prime+1;n}(\beta_{l^\prime,l })+ (1+\mu_l)K_{n^\prime;n}(\beta_{l^\prime,l })+\gamma_lK_{n^\prime-1;n }(\beta_{l^\prime,l })] \hat{a}^\text{in}_{n^\prime,l^\prime,p^\prime}\nonumber\\
      \hat{a}^\text{out}_{n, l,  p }(\omega^\text{out})=-\sum_{n^\prime} [\nu_lK_{n^\prime+1;n }(\beta_{\Delta l^*})+ (1+\mu_l)K_{n^\prime;n }(\beta_{\Delta l^*})+\gamma_lK_{n^\prime-1;n}(\beta_{\Delta l^*})] \hat{a}^\text{in}_{n^\prime,l +\Delta l^*, p }\nonumber\\
\end{align}
choosing $ \omega^\text{out}=\omega^\text{in}-\Delta l^* \Omega_0$ then $\beta=0$ so $ K_{n^\prime,n }=\delta_{n^\prime,n }$
\begin{align}
      \hat{a}^\text{out}_{n ,l,  p}=-\nu_l\hat{a}^\text{in}_{n -1,l +\Delta l^* ,p }- (1+\mu_l)\hat{a}^\text{in}_{n,l +\Delta l^*, p }-\gamma_l\hat{a}^\text{in}_{n+1,l +\Delta l^*, p}\nonumber\\
\end{align} 
The proposed input state for the non-classical strategy is
\begin{align}
|\Psi_\text{Q}\rangle^\text{in}=\hat{D}_{n +1,l +\Delta l^*, p }(\alpha_\text{Q})\hat{S}_{n ,l+\Delta l^*, p }(\xi=s)|0\rangle,
    \label{F4}
\end{align}
where $s\in\reals$, $\alpha_Q=i|\alpha_Q| $ where $|\alpha_\text{Q}|^2 = N^\text{Coh}$ (and $N= N^\text{Coh} + N^\text{Sq} $).\\
So the mean displacement and the covariance matrix for the state to be measured, $\vec{m} = \lbrace n, l, p \rbrace $ are
\begin{align*}
\Sigma_{\mathbf{m}} 
&=(1-\eta)\begin{pmatrix}
e^{-2s}+|\mu|^2e^{2s}  &  |\mu|(e^{-2s}-e^{2s})   \\
 |\mu|(e^{-2s}-e^{2s}) & e^{2s}+|\mu|^2e^{-2s}
\end{pmatrix}+\begin{pmatrix}
(1-\eta)(|\nu|^2+|\gamma|^2)+\eta  &  0   \\
 0 & (1-\eta)(|\nu|^2+|\gamma|^2)+\eta
\end{pmatrix},\\
\mathbf{r}_{\mathbf{m}}  &=
\sqrt{1-\eta}\begin{pmatrix}
2|\gamma||\alpha_Q| \\
0
\end{pmatrix}.
\end{align*}
Let's compute the relevant derivatives,
\begin{align*}
\partial_\Omega(\mu) =\tau\Delta l^* \quad \quad
\partial_\Omega(\gamma) = \frac{\sqrt{2}}{\sigma}\sqrt{\frac{n+1}{2}}\Delta l^*.
\end{align*}
Considering homodyne in the position quadrature, $\bar{q}=\mathbf{r}_{\mathbf{m}}^{x} $ and variance $\Sigma_q= \Sigma_{\mathbf{m}}^{xx}$. The Fisher information \eqref{CFisher}, is
\begin{align}
    F_\text{Q} = \frac{4(1-\eta) N^\text{Coh} \frac{2}{\sigma^2}\frac{n+1}{2}(\Delta l^*)^2  }{(1-\eta) e^{-2s} +\eta}.
\end{align}
In the limit of large squeezing $e^{2 s}\approx 2 N^\text{Sq} $ and in the noiseless scenario $\eta=0 $ the Fisher information reduces to
\begin{align}
   F_{\text{Q}} \xrightarrow[\eta=0,\; N^{\text{Sq}}\gg 1]{}
\frac{4 N^{\text{Coh}} \frac{n+1}{\sigma^2} (\Delta l^*)^2}{\frac{1}{2N^{\text{Sq}}}}=8 N^\text{Coh}N^\text{Sq} \frac{n+1}{\sigma^2}(\Delta l^*)^2  
\label{E6}
\end{align}
Given a fixed energy and therefore a fixed number of photons $N=N^\text{Coh}+N^\text{Sq}$ it is optimal to allocate the energy $ N/2=N^\text{Coh}=N^\text{Sq}$, then the Fisher information \eqref{E6} can be rewritten as
\begin{align}
    F_{\text{Q}} \xrightarrow[\eta=0,\; N^{\text{Sq}}\gg 1]{} 2 N^2 \frac{n+1}{\sigma^2}(\Delta l^*)^2 . 
\end{align}
It is clear then that in the noiseless scenario the presented strategy follows the Heisenberg scaling asymptotically.

Now we analyze the classical strategy. The input state is
\begin{align}
    |\Psi_\text{C}\rangle^\text{in}:=\hat{D}_{n +1,l +\Delta l, p}(\alpha_\text{C})|0\rangle,
\end{align}
with $|\alpha_\text{C}|^2=N$.
The corresponding Fisher information is
\begin{align}
    F_\text{C}=4(1-\eta)N|U_{\mathbf{m}\mathbf{l}}'|^2=4(1-\eta)N|\partial_\Omega \gamma|^2=4(1-\eta) N \frac{2}{\sigma^2}\frac{n+1}{2}(\Delta l^*)^2.
\end{align}
We are interested in determining if there is any advantage to consider the squeezed state \eqref{F4} over a classical strategy that only consist in displacement for noisy scenarios. To do so we consider the ratio introduced in the main text:
\begin{align}
   \frac{F_\text{Q}}{F_\text{C}} =\frac{N^\text{Coh}}{N} \frac{1}{(1-\eta)\left( 2N^\text{Sq} + 1 - 2\sqrt{ N^\text{Sq} \left( N^\text{Sq} + 1 \right) }\right) +\eta}.
\end{align}

The maximization of this ratio constrained by the condition $N = N^\text{Coh}+ N^\text{Sq}$ is defined as
\begin{equation}
R=\max_{N^\text{Sq},N^\text{Coh}}   \frac{F_\text{Q}}{F_\text{C}}  \quad \textrm{s.t.} \quad N=N^\text{Sq}+N^\text{Coh},
\end{equation}
For the metasurface we find that
\begin{equation}
    R= \frac{1+2N\eta - \sqrt{1+4N(1-\eta)\eta}}{2N\eta^2},
\end{equation}
constrained by the following function $N^\text{Coh} = N^\text{Coh}(N, \eta) \geq 0$:
\begin{equation}
   N^\text{Coh}(N, \eta) =  \frac{2 N \left(\eta ^2 (4 N+2)-\eta  \left(\sqrt{1-4 (\eta -1) \eta  N}+4 N+2\right)+\sqrt{1-4 (\eta -1) \eta  N}-1\right)+\sqrt{1-4 (\eta -1) \eta  N}-1}{2 \eta  (\eta +4 (\eta -1) N-2)},
\end{equation}
which for the noiseless limit $\eta \rightarrow 0  $ reduces to $N^\text{Coh} = N(1+N)/(1+2N)$.

\subsection{Complex surface}\label{app.ComplexSurface}
In the Appendix~\ref{Roughmodel} we present a model for a rough surface. In the following we compute the Fisher information for the state considered in the main text. The transformation is,
\begin{align}
      \hat{a}^\text{out}_{n,l , p }(\omega^\text{out})&=\sum_{l^\prime p^\prime n^\prime} c_{l^\prime,p^\prime;l,  p} [\nu_lK_{n^\prime+1;n }(\beta_{l^\prime,l })+ (1+\mu_l)K_{n^\prime;n }(\beta_{l^\prime,l })+\gamma_lK_{n^\prime-1;n }(\beta_{l^\prime,l })] \hat{a}^\text{in}_{n^\prime,l^\prime,p^\prime}\nonumber\\ &\quad\quad\quad\quad\quad \quad\quad\quad \beta_{l,l^\prime}:=\frac{\omega^\text{in}-\Delta l\Omega-\omega^{\text{out}}}{\sigma}\,
\end{align}
where 
\begin{align}
    c_{l^\prime,p^\prime;l , p } = (-\sqrt{1-\varepsilon^2}+i\frac{\varepsilon}{\mathcal{N}})\:\delta_{l^\prime,l }\delta_{p^\prime, p }-i\varepsilon \frac{ e^{-\frac{(l^\prime-l)^2}{4\sigma_l^2}} e^{-\frac{(p^\prime- p )^2}{4\sigma_p^2}}}{\mathcal{N}}e^{i\theta_{l^\prime ,p^\prime;l, p}} .
\end{align}
For the proposed state in the main text,
\begin{align*}
|\Psi_\text{Q}\rangle^\text{in}=\hat{D}_{n +1,l +\Delta l, p}(\alpha_\text{Q})\hat{S}_{n ,l , p}|0\rangle
\end{align*}
where $\alpha_Q=|\alpha_Q|e^{-i(\theta_{l, p;l+\Delta l, p})}$ and $\Delta l$ is now a fixed value determined by the probe state. The displacement and covariance matrix after the transformation are,
\begin{align*}
\Sigma_{\mathbf{m}} 
&=(1-\varepsilon^2)(1-\eta)\begin{pmatrix}
e^{-2s}+|\mu|^2e^{2s}  &  |\mu|(e^{-2s}-e^{2s})   \\
 |\mu|(e^{-2s}-e^{2s}) & e^{2s}+|\mu|^2e^{-2s}
\end{pmatrix}+[(1-\eta)(\varepsilon^2+(1-\varepsilon^2)(|\nu^2|+|\gamma^2|))+\eta]\begin{pmatrix}
1  &  0   \\
 0 & 1
\end{pmatrix},\\
\mathbf{r}_{\mathbf{m}}  &=
\begin{pmatrix}
2|U_{\mathbf{m}\mathbf{l}}||\alpha| \\
0
\end{pmatrix}=\begin{pmatrix}
2|\varepsilon\frac{ e^{-\frac{(\Delta l)^2}{4\sigma_l^2}} }{\mathcal{N}}\nu||\alpha_Q||\sqrt{1-\eta}| \\
0
\end{pmatrix},
\end{align*}
where we made the choice $ \omega^\text{out}=\omega^\text{in}-\Delta l \Omega_0$ and used that for $\Omega\ll \sigma$ we can approximate $K_{n;m}\approx \delta_{n,m}$  Then, the Fisher information after homodyne in the position quadrature is \eqref{CFisher} is, 
\begin{equation}
    F_\text{Q} =\frac{[2\varepsilon \frac{e^{-\frac{(\Delta l)^2}{4\sigma_l^2}}}{\mathcal{N}}  \Delta l\frac{n+1}{\sigma} ]^2|\alpha_\text{Q}|^2(1-\eta)}{(1-\eta)(1-\varepsilon^2)e^{-2s_\mathbf{i}}+(1-\eta)\varepsilon^2+\eta}
       \label{eq:E13}
\end{equation}
On the other hand the Fisher information of the classical strategy given by the input state,
\begin{align}
|\Psi_\text{C}\rangle^\text{in}=\hat{D}_{n +1,l +\Delta l, p }(\alpha_\text{C})|0\rangle.
\end{align}
is
\begin{equation}
    F_\text{C} = [2\varepsilon \frac{e^{-\frac{(\Delta l)^2}{4\sigma_l^2}}}{\mathcal{N}}  \Delta l\frac{n+1}{\sigma} ]^2|\alpha_\text{C}|^2(1-\eta)
    \label{eq:E15}.
\end{equation}
When computing the ratio, the dependence with the parameters of the beam disappear. Writing the squeezing and the displacement in terms of the number of photons the ratio is, that is, using the fact that $|\alpha_\text{C}|^2 = N$, $|\alpha_\text{Q}|^2 = N^\text{Coh}$ and $N = N^\text{Coh} + N^\text{Sq}$, we find
\begin{align*}
    R=\frac{F_\text{Q}}{F_\text{C}}=\frac{N^\text{Coh}}{N}\frac{1}{(1-\eta)(1-\varepsilon^2)\left( 2N^\text{Sq} + 1 - 2\sqrt{ N^\text{Sq} \left( N^\text{Sq} + 1 \right) }\right)+(1-\eta)\varepsilon^2+\eta}.
\end{align*}

\end{section}

\section{Benchmarking with the quantum Fisher information}\label{sec:appG}
We compute the QFI of the quantum strategy for both, metasurface and complex surface.
\begin{align}
   J = \frac{1}{2} \frac{\text{Tr}[(\Sigma_{\mathbf{m}}^{-1}\Sigma_{\mathbf{m}}')^2]}{1+P} + (\mathbf{r}_{\mathbf{m}}') ^{\trans}\Sigma_{\mathbf{m}}^{-1}\mathbf{r}'_{\mathbf{m}}\quad \text{ where } \quad P= \frac{1}{\sqrt{\det(\Sigma_{\mathbf{m}})}}.
\end{align}
After substituting and simplifying, for the metasurface 
\begin{align*}
    J_\text{meta}=\frac{
4(1+n)|\alpha_\text{Q}|^{2}(\Delta l^*)^{2}\,(1-\eta)}{\sigma^{2}\Bigl[e^{-2s}(1-\eta)+\eta\Bigr]} +\frac{
\bigl(e^{-2s}-e^{2s}\bigr)^{2}
(\Delta l^*)^{2}
(1-\eta)^{2}
\tau^{2}
}{
\Bigl[
e^{-2s}(1-\eta)+\eta
\Bigr]
\Bigl[
e^{2s}(1-\eta)+\eta
\Bigr]
\left(
1+\frac{1}{\sqrt{\mathcal{D}}}
\right)
}
\end{align*}
where $\mathcal{D}=1 - 2\eta+ e^{-2s}\eta+ e^{2s}\eta+ 2\eta^{2}- e^{-2s}\eta^{2}-e^{2s}\eta^{2}$.\\

We can obtain an expression for the ratio between the QFI and the classical Fisher information of our quantum protocol $F_\text{Q}$, which was given by
\begin{equation}
    F_\text{Q} = \frac{4(1-\eta) N^\text{Coh} \frac{2}{\sigma^2}\frac{n+1}{2}(\Delta l^*)^2  }{(1-\eta) e^{-2s_\mathbf{i}} +\eta}.  
\end{equation}
We find
\begin{align}
    &\frac{J_\text{meta}}{F_\text{Q}}   = \\
&\frac{
e^{s} 
\left(
(1 - \eta+2 N_{\mathrm{Sq}} - 2 \sqrt{N_{\mathrm{Sq}} (1 + N_{\mathrm{Sq}})})
(1 - \eta) 
\right)
\sigma^{2}
\left(
\frac{4 (1+n)\,\alpha^{2}}{\sigma^{2}}
+
\frac{4 (1-\eta)\,\tau^{2} \sinh^{2}(2s)}
{\left(e^{2s} (1-\eta) + \eta\right)\left(
1 + \frac{1}{\sqrt{
1 - 2 (1-\eta)\eta
+2 (1-\eta)\eta \cosh(2s)
}}
\right)}
\right)
}{
4 (1+n)\, N_{\mathrm{coh}} 
\left(
\cosh s + (-1+2\eta)\sinh s
\right)
}.
\end{align}

The noiseless limit of the metasurface's QFI is 
\begin{equation}
    \lim_{\eta \rightarrow 0} J_\text{meta} \equiv J^{(0)}_\text{meta}=(\Delta l^*)^2 \left(
\frac{2 e^{2s} (1 + n)\,|\alpha_\text{Q}|^2}{\sigma^2}
+ \tau^2 \sinh^2(2s)
\right)
\end{equation}

This is to be compared to the noiseless version of our $F_\text{Q}$ (removing the index $\vec{i}$):
\begin{equation}
   \lim_{\eta \rightarrow 0} F_\text{Q} \equiv F^{(0)}_\text{Q} = 4 e^{2s} N^\text{Coh} \frac{n+1} {\sigma^2}(\Delta l^*)^2 .
\end{equation}

This noiseless ratio is found to be:
\begin{equation}
   \frac{J^{(0)}_\text{meta}}{F^{(0)}_\text{Q}}   = \frac{
\left(
1 + 2 N_{\mathrm{Sq}} - 2 \sqrt{N_{\mathrm{Sq}} (1 + N_{\mathrm{Sq}})}
\right)
\left(
2 e^{2s} (1+n)\,|\alpha_\text{Q}|^{2}
+ \sigma^{2}\tau^{2}\sinh^{2}(2s)
\right)
}{
2 (1+n)\, N_{\mathrm{coh}}
}.
\end{equation}
Inserting $e^{-2s} = 1 + 2 N_{\mathrm{Sq}} - 2 \sqrt{N_{\mathrm{Sq}} (1 + N_{\mathrm{Sq}})} $, $\sinh^2s  = N_{\mathrm{Sq}}$, $\sinh (2s) = 2\sqrt{N_{\mathrm{Sq}}(1+N_{\mathrm{Sq}})}$ and $|\alpha_\text{Q}|^2 = N_{\mathrm{coh}}$ we simplify the expression to

\begin{equation}
   \frac{J^{(0)}_\text{meta}}{F^{(0)}_\text{Q}}   = \frac{
\left(
1 + 2 N_{\mathrm{Sq}} - 2 \sqrt{N_{\mathrm{Sq}} (1 + N_{\mathrm{Sq}})}
\right)
\left(
 \frac{(1+n)\,N_{\mathrm{coh}} }{1 + 2 N_{\mathrm{Sq}} - 2 \sqrt{N_{\mathrm{Sq}} (1 + N_{\mathrm{Sq}})}}
+ 2\sigma^{2}\tau^{2}N_{\mathrm{Sq}} (1 + N_{\mathrm{Sq}})
\right)
}{
 (1+n)\, N_{\mathrm{coh}}
}.
\end{equation}
Note that because both strategies make use of the same probe state, their difference lies solely on the measurement: $F^{(0)}_\text{Q}$ is limited by homodyne detection, while $J^{(0)}_\text{meta}$ is by definition optmized over all POVMs. The ratio then gives us information about the regions where homodyne is less optimal. This trade-off is non-trivial depending on the numeric values of parameters $\tau$ and $\sigma$, as well as in the mode number $n$. Fixing these, one would need to then weight the practical cost of implementing an optimal POVM against the simplicity of homodyne detection depending on the resulting factor between both strategies. A straightforward estimation shows, however, that the advantage of an optimal POVM over homodyning could be important for bright states, a region which could be in conflict with the no measurement back-action condition, imposed in the main text and fundamental for obtaining an i.i.d. setting in agreement with the asymptotic analysis that underlies the formalism of the Fisher information. Further analysis of how an optimal POVM deviates from standard detection schemes such as photon-counting and homodyne is left for future work.

For the complex surface the QFI is
\begin{align*}
J_\text{cs}=\frac{\Delta l^{2}(1-\eta)}{-\varepsilon^{2}(1-\eta)+ e^{-2s}(1-\varepsilon^{2})(1-\eta)+ \eta}\left[\frac{4 e^{\frac{\Delta l^{2}}{2\sigma_l}}\,n\,\alpha^{2}\,\varepsilon^{2}}{\mathcal{N}^{2}\sigma^{2}}-\frac{
e^{-4s}\bigl(-1+e^{4s}\bigr)^{2}(1-\varepsilon^{2})^{2}(1-\eta)\sqrt{\mathcal{D}}\tau^{2}}{\Bigl[-\varepsilon^{2}(1-\eta)+ e^{2s}(1-\varepsilon^{2})(1-\eta)+ \eta\Bigr]\left(1+\sqrt{\mathcal{D}}\right)}\right]
\end{align*}
where 
\begin{align*}
\mathcal{D} =&1- e^{-2s}(1-\varepsilon^{2})\bigl[\varepsilon^{2}(1-\eta)-\eta\bigr](1-\eta)- e^{2s}(1-\varepsilon^{2})\bigl[\varepsilon^{2}(1-\eta)-\eta\bigr](1-\eta) \\
& + 2\varepsilon^{4}(1-\eta)^{2}- 2\eta+ 2\eta^{2}+\varepsilon^{2}\bigl(-2+6\eta-4\eta^{2}\bigr).
\end{align*}
And in the noiseless regime reduces to
\begin{align*}
J_\text{cs}^{(0)}=\frac{\Delta l^{2}}{-\varepsilon^{2}+ e^{-2s}(1-\varepsilon^{2})}\left[\frac{4 e^{\frac{\Delta l^{2}}{2\sigma_l}}\,n\,\alpha^{2}\,\varepsilon^{2}}{\mathcal{N}^{2}\sigma^{2}}-\frac{
e^{-4s}\bigl(-1+e^{4s}\bigr)^{2}(1-\varepsilon^{2})^{2}\sqrt{\mathcal{D}}\tau^{2}}{\Bigl[-\varepsilon^{2}+ e^{2s}(1-\varepsilon^{2})\Bigr]\left(1+\sqrt{\mathcal{D}}\right)}\right]
\end{align*}
where 
\begin{align*}
\mathcal{D} =&1- e^{-2s}(1-\varepsilon^{2})\varepsilon^{2} - e^{2s}(1-\varepsilon^{2})\varepsilon^{2} 
 + 2\varepsilon^{4}-2 \varepsilon^{2}\bigl.
\end{align*}
The previous expressions are too complex to draw general conclusions about the influence of the probe's parameters in the difference between the homodyne Fisher information and the QFI. This information should be computed in each particular situation.

\end{document}